\documentclass[12pt,preprint]{aastex}

\usepackage{psfig}

\shorttitle{The Nuclear Gas Dynamics and Star Formation of NGC\,7469}
\shortauthors{}

\newcommand{\kms}{\,\hbox{\hbox{km}\,\hbox{s}$^{-1}$}}

\begin{document}

\title{The Nuclear Gas Dynamics and Star Formation of 
NGC\,7469\footnote{The near infrared data presented herein were
obtained at the W.M. Keck Observatory, which is operated as a
scientific partnership among the California Institute of Technology,
the University of California and the National Aeronautics and Space
Administration. The Observatory was made possible by the generous
financial support of the W.M. Keck Foundation.
The radio data are based on observations carried out with the IRAM
Plateau de Bure Interferometer. 
IRAM is supported by INSU/CNRS (France), MPG 
(Germany) and IGN (Spain).}}

\author{R.I. Davies, L.J. Tacconi, and R. Genzel}
\affil{Max-Planck-Institut f\"ur extraterrestrische Physik, 
Postfach 1312, 85741, Garching, Germany}

\begin{abstract}
We report interferometric radio CO\,2-1 and HCN\,1-0 observations at
resolutions of 0.7\arcsec\ and 2.0\arcsec\ respectively, and
0.085\arcsec\ resolution adaptive optics K-band spectroscopy,
including H$_2$ 1-0\,S(1) line emission and CO\,2-0 stellar
absorption, of the inner few arcseconds of NGC\,7469.
The CO\,2-1 map shows a ring of molecular clouds (which in general lie
outside the compact knots seen in K-band images) and a bright extended
nucleus, with a bar or pair of spiral arms between them.
The dynamical structure of both the radio CO\,2-1
and the K-band H$_2$ 1-0\,S(1) lines at their different resolutions can
be reproduced using
a single axisymmetric mass model comprising 3 components: a broad
disk, a ring 4--5\arcsec\ across, and an extended nucleus which we
interpret as an inner nuclear ring about 0.5\arcsec\ across.
The velocity residuals between the model and the data have a standard
deviation of 25\kms, and no non-circular motions faster than this are
seen, although this may be because in some cases a secondary bar is
not expected to cause gas inflow.
From the dynamical mass and estimates of the stellar mass we find
that the CO-to-H$_2$ conversion is 0.4--0.8 times that for the Milky
Way, following the trend to small factors that has been previously
reported for intense star forming environments.
The central H$_2$ 1-0\,S(1) morphology has a strong peak at the
nucleus, but this does not trace the mass distribution;
the rotation curves indicate that there is no strong nuclear mass
concentration.
The origins of the 1-0\,S(1) emission are instead likely to lie in
X-ray and UV irradiation of gas by the AGN rather than via processes
associated with star formation.
Using the 2.3\,$\mu$m stellar CO\,2-0 bandhead absorption and the slope
of the continuum we have directly resolved the nuclear star cluster to
be 0.15--0.20\arcsec\ across, and find that it is asymmetric.
This cluster has an age of less than about 60\,Myr and contributes
20--30\% of the nuclear K-band light, and about 10\% of the nuclear
bolometric luminosity.
Within a radius of $\sim$4\arcsec\ gas contributes more than half
the total mass;
but in the nucleus, within a radius of 0.1\arcsec, it is likely that 
most of the mass is due instead to stars.
\end{abstract}


\section{Introduction}
\label{sec:intro}

The SBa Seyfert 1 galaxy NGC\,7469 is a luminous infrared source with
$L_{\rm bol}\sim3\times10^{11}$\,L$_\odot$, 
assuming a distance of 66\,Mpc (taking
H$_0=75$\kms\,Mpc$^{-1}$ and $V_{\rm LSR}=4925$\kms, \citealt{mei90}).
Much of the interest in the galaxy has been focussed on the 
circumnuclear ring structure on scales of 1.5--2.5\arcsec, which has
been observed at 
radio \citep{wil91,con91,col01},
optical \citep{mau94}, 
mid infrared \citep{mil94,soi03}, 
and near infrared \citep{gen95,lai99,sco00} 
wavelengths.
These data suggest that recent star formation in this
ring contributes more than half of the entire bolometric luminosity of
the galaxy.
Additionally, as much as one third of the K-band continuum within
1\arcsec\ of the nucleus may also
originate in stellar processes rather than the AGN itself
\citep{maz94,gen95}.
NGC\,7469 is therefore a key object for studying the relation between
circumnuclear star formation and an AGN, and how gas is brought in to
the nucleus to fuel these processes.

This crucial question is still the subject of much debate.
It has been known for some time that bars can form in the disks
of galaxies, and that the shocks in the diffuse gas associated with
these can drive the gas from the disk on scales of 10\,kpc down to the
nucleus on scales of 1\,kpc.
However, it is only more recently that the issue of whether or not bars 
within bars might be able to drive the gas even further to the centre
has been studied in more detail, both observationally
\citep[e.g.][]{mar99,reg99b,lai02}
and theoretically \citep[e.g.][]{mac02b,shl02}.
One classic example is the Seyfert 2 NGC\,1068 which has a
circumnuclear ring at 15\arcsec\ (1\,kpc), inside of which is a bar
and near the nucleus another gaseous ring at 1\arcsec\ (70\,pc)
\citep{hel97,tac97,sch00,bak00}.
We wish to investigate whether dynamical processes may be
operating in NGC\,7469 which help to drive gas from the circumnuclear ring
further towards the centre.

To do so we have brought together the unique combination of high
resolution radio CO data and near infrared adaptive optics 
H$_2$ 1-0\,S(1) data, giving us a 
tool with which we can probe the distribution and dynamical structure
of the 
molecular gas across nearly 2 orders of magnitude in spatial scale.
Our data are presented in Section~\ref{sec:obs}.
In Section~\ref{sec:CO} we consider the properties of the molecular gas
with respect to both the radio CO\,2-1 and near infrared 1-0\,S(1)
emission.
We present a single mass model which is able to reproduce the
kinematics of both sets of data, and discuss the issue of
non-circular motions associated with bars and spiral arms.
Finally in this section, we turn to the excitation process of the
1-0\,S(1) emission, and the CO-to-H$_2$ conversion ratio.
Section~\ref{sec:starclus} discusses the nuclear star formation.
Finally, our conclusions are summarised in Section~\ref{sec:conc}.

\section{Observations and Data Reduction}
\label{sec:obs}

\subsection{Radio Data}
\label{sec:obs1}

We have mapped the $^{12}$CO\,J=2$\rightarrow$1 and
HCN\,J=1$\rightarrow$0 lines in NGC\,7469 with the IRAM millimeter 
interferometer on the Plateau de Bure, France \citep{gui92}.
The CO\,2-1 and HCN\,1-0 data were obtained in January 1997.
The array consisted of 5 15-m antennas positioned in 2 
configurations providing 20 baselines ranging from 32--400\,m. 
We observed NGC\,7469 for $\sim$9\,hrs in each configuration for a
total of 18\,hrs of integration time.
The antennas were equipped with 
dual frequency SIS receivers enabling us to observe the CO and HCN lines
simultaneously.
SSB system temperatures above the atmosphere were $\sim$120\,K at the
86\,GHz frequency of the HCN\,1-0 line and $\sim$350\,K at the
228\,GHz frequency of the CO\,2-1 line.
3C345.3 was observed for bandpass calibration as well as every half
hour for phase and amplitude calibration.
Absolute fluxes were established by assigning flux of 3.5\,Jy  and 6.0\,Jy
to 3C454.3 at 228\,GHz and 86\,GHz respectively.  Fluxes of strong sources are 
determined through careful 
monitoring with both the interferometer and the IRAM 30-meter telescope 
on Pico Veleta, Spain. 
Based on the monitoring measurements, we assume 
that the accuracy of the flux scale is better than 20\%.
The phase noise on the longest baselines was 20$^\circ$--30$^\circ$,
even at the higher frequency. 
Spectral resolution of 2.5\,MHz, corresponding to
$\sim$6.5\kms\ for the HCN\,1-0 line and
3.3\kms\ for the CO\,2-1 line, was provided by 6
autocorrelator spectrometers covering the total receiver bandwidth of
500\,MHz (1300\kms\ for HCN and 650\kms\ for CO\,2-1).

In an earlier set of runs between October and December 1994, the
$^{12}$CO\,J=1$\rightarrow$0 line was mapped using 4 antennae in 4
configurations providing 24 baselines in the range 24--288\,m.
The data were calibrated and reduced in the same way, yielding a final
resolution of 2.5\arcsec.
Because the CO\,2-1 data supercede these data, we use the older data only in
Section~\ref{sec:CO-H2} to estimate the CO-to-H$_2$ conversion factor.
The map itself has been presented previously by \citet{tac96}.

All the data were first calibrated using the IRAM CLIC software.
We then made uniformly weighted channel maps for the CO\,2-1 and HCN
data.
We CLEANed all the maps using software available as part of the GILDAS 
package.
To increase the sensitivity maps were made with a velocity 
resolution of 20\kms\ in all three lines.
The CLEANed maps were reconvolved with 0.7\arcsec and 2\arcsec\ FWHM
Gaussian beams for the CO\,2-1 and HCN\,1-0 lines respectively.
The rms noise after CLEANing is 1\,mJy\,beam$^{-1}$ in the HCN\,1-0
maps, and 4\,mJy\,beam$^{-1}$ in the higher resolution CO\,2-1 maps.

\begin{figure}
\centerline{\psfig{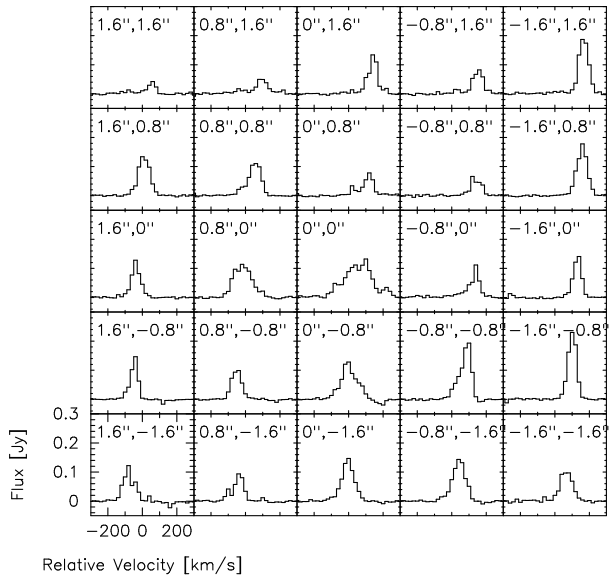}}
\caption{Sample spectra from the central 3\arcsec$\times$3\arcsec\ of
the CO\,2-1 map. 
The spectral resolution is 20\kms.
The position of each spectrum is indicated in the upper left corner of
each box, as an offset relative to the nuclear pointing.
The velocity scale is relative to the `nominal' systemic velocity of
4925\kms.
}
\label{fig:COspectra}
\end{figure}

The CO is centered kinematically at $V_{\rm LSR} = 4925$\kms,
consistent with that found by \cite{mei90} in their CO observations.
The integrated CO\,2-1 and HCN\,1-0 line maps are shown in
Fig.~\ref{fig:COmap}, while sample spectra from the central 3\arcsec\
of the CO\,2-1 channel maps are shown in Fig.~\ref{fig:COspectra}.
From this figure it can be seen that, even at sub-arcsecond resolution,  
we have achieved a signal-to-noise ratio of at least 20 throughout the 
central regions of NGC\,7469.

\subsection{Near Infrared Data}
\label{sec:obs2}

\begin{figure}
\centerline{\psfig{file=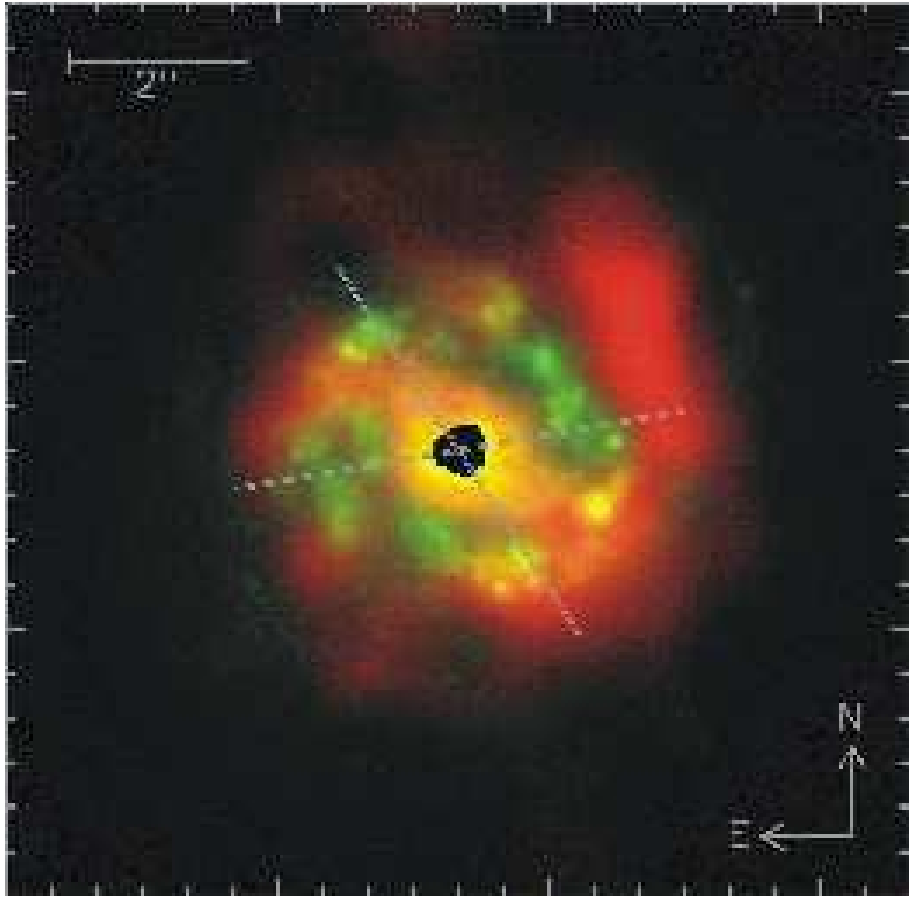,width=12cm}}
\caption{Overlay showing the stellar (J-band) light in green on the
gas (CO\,2-1) distribution in red; regions with both are yellower.
The two images have been aligned assuming the nuclei are centered
in the same position.
The ring of star forming knots lies mostly inside the ring of gas clouds.
The J-band has been used rather than the K-band because the central
PSF is less of a problem, although the central region where the
nuclear light is strong has still been masked out.
In both the J and K-bands the star forming knots are the same.
The J-band {\em HST} image was kindly provided by A.~Evans.
The positions and widths of the slits are used for the K-band
spectroscopy are drawn in grey.
}
\label{fig:StarsGas}
\end{figure}

\begin{figure}
\centerline{\psfig{file=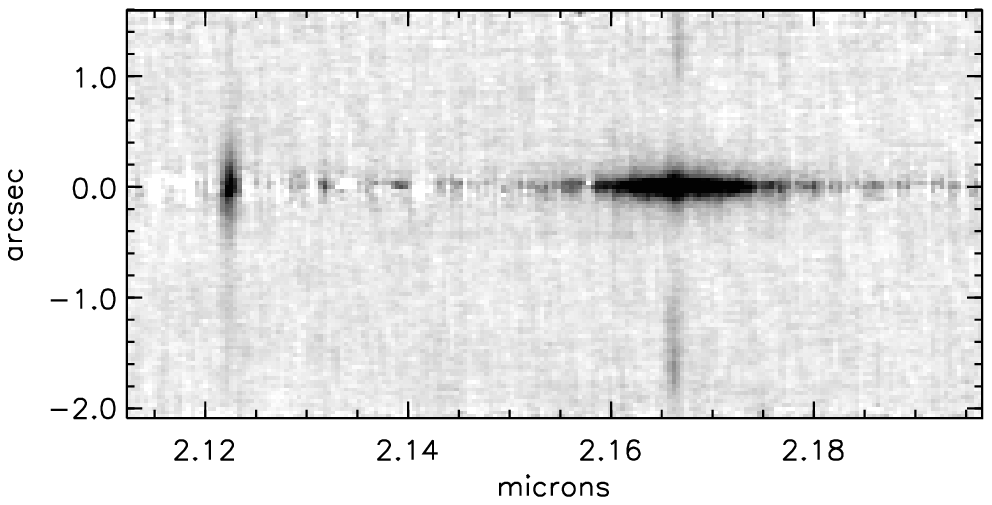,width=8cm}\hspace{5mm}\psfig{file=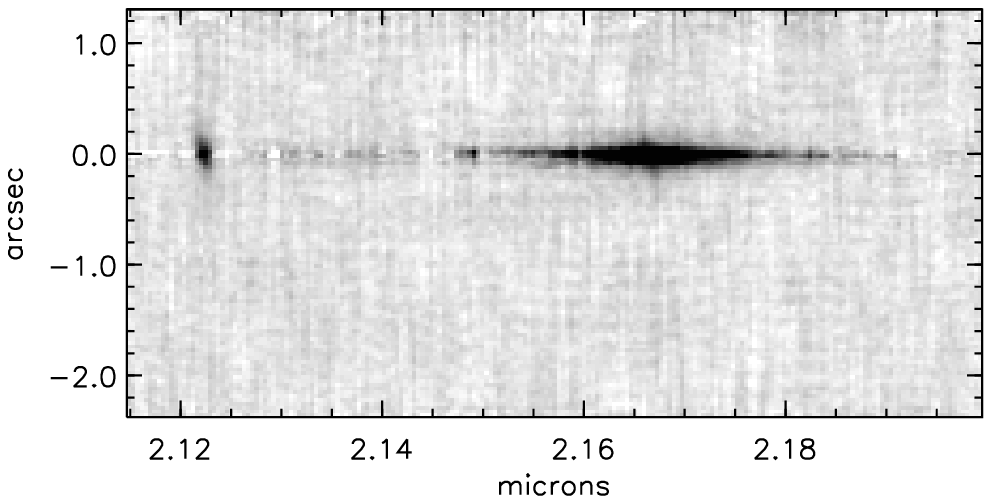,width=8cm}}
\caption{Longslit spectra of NGC\,7469 at position angles 
33$^\circ$ (left) and 100$^\circ$ (right).
A cubic spline fit to each row (using the regions free of emission
lines) has been subtracted.
The narrow 1-0\,S(1) line is to the left; the broad and narrow
components of the Br$\gamma$ are to the right.
Velocity gradients across the nucleus can clearly be seen.
}
\label{fig:2dspec}
\end{figure}

The data were obtained on the nights of 11--12 Nov 2002 using NIRSPEC
behind the adaptive optics system on the Keck II telescope.
Spectra were obtained in low resolution mode in the K-band, providing
a nominal resolution of R=2000 with a slit 37\,mas wide and
3.93\arcsec\ long.
The wavelength coverage is sufficient to include all the K-band in a
single setting.
Two position angles were used: 33$^\circ$ (to bisect two of the bright
star forming knots in the K-band image of \citealt{sco00}) and
100$^\circ$.
The latter was chosen to include the supernova remnant \citep{col01},
although this proved unsuccessful due to uncertainties in the
positioning and the narrowness of the slit.
The positions and widths of the slits are shown in
Fig.~\ref{fig:StarsGas} on top of a composite of the {\em Hubble Space
Telescope} (HST) J-band image (green) and CO image (red).
Individual exposures of 5\,min were used, and the object was moved to a
new pixel row after each exposure. 
In total integration times 85\,min and 100\,min were achieved at the
two position angles respectively.
Since the slit was not long enough to permit nodding, a few sky
frames were taken.
Standard calibrations were performed, including atmospheric standard
stars (using the type B9\,V star HD\,218624), arcs, flatfields, and dark
frames.
The data were reduced using PC-IRAF 2.11.3 using standard techniques,
with the single exception that all the frames were resampled at half
the pixel scale before applying any rotation or curvature correction.
The final step was to rebin the data once all the exposures were
combined, providing a spectral scale of $4.287\times10^{-4}$\micron\
and a spatial scale of 34.8\,mas.
The resulting spectral resolution, as measured from the arc lamp lines,
was 104\kms\ FWHM.
Continuum subtracted images of the spectral region covering the
1-0\,S(1) and Br$\gamma$ lines are shown in Fig.~\ref{fig:2dspec}.

The reference for the adaptive optics system was NGC\,7469 itself.
The correction provided -- and hence spatial resolution achieved -- by
the AO was similar on both nights due to the excellent
atmospheric conditions.
It was measured on the second night by taking a spectrum of a star,
after using neutral 
density filters to reduce its count rate on the wavefront sensor to
that measured on NGC\,7469 itself.
Additionally, the AO parameters (frame rate and gains) were set to
those used on the object.
The spectrum was reduced in exactly the same way.
As expected, the resolution improves at longer wavelengths, but is
measured to be 85\,mas near the 1-0\,S(1) and Br$\gamma$ lines, and
80\,mas near the CO absorption at 2.3\micron.
An alternative method to estimate the spatial resolution directly from
the AGN itself is described in the text.

Flux calibration using the atmospheric standard star yielded K-band
magnitudes of 13.97 and 14.01 within a 0.5\arcsec\ length of the
37\,mas slit, for position angles 33$^\circ$ and 100$^\circ$
respectively.
Extrapolating the profile symmetrically over a circular aperture with
diameter 0.5\arcsec\ gives $K=12.0$; 
similarly in a 1\arcsec\ aperture we find $K=11.7$.
However, we note that there is a considerable uncertainty
associated with the calibration due to the difficulty of positioning
both the standard star (with a diffraction limited FWHM of about
45\,mas) and NGC\,7469 (which had poorer correction) on a slit
which is only 37\,mas wide.

\section{Molecular Gas}
\label{sec:CO}

\subsection{Distribution and Properties of the CO Emission}
\label{sec:COprop}

\begin{figure}
\centerline{\psfig{file=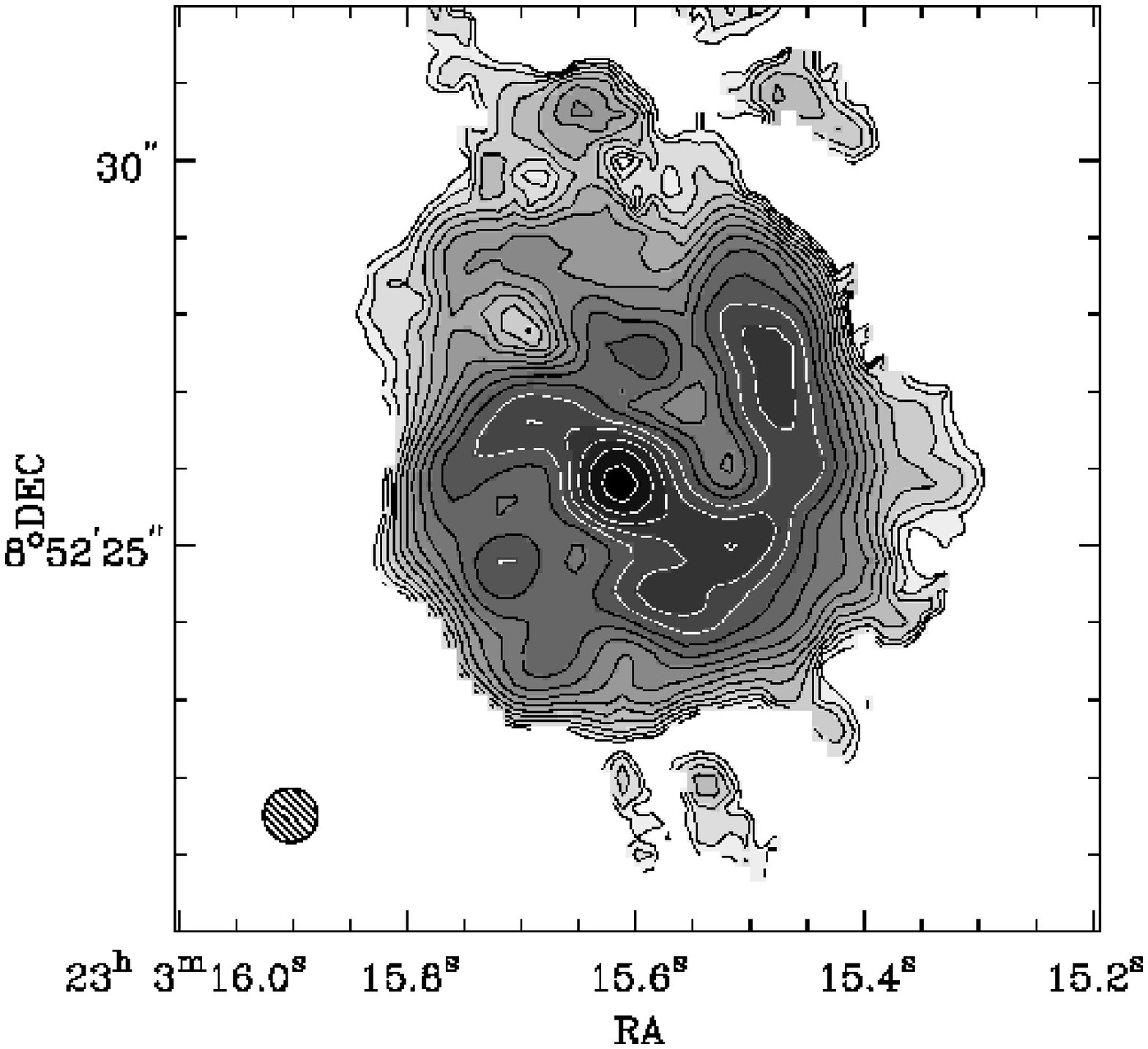,width=8cm}\hspace{10mm}\psfig{file=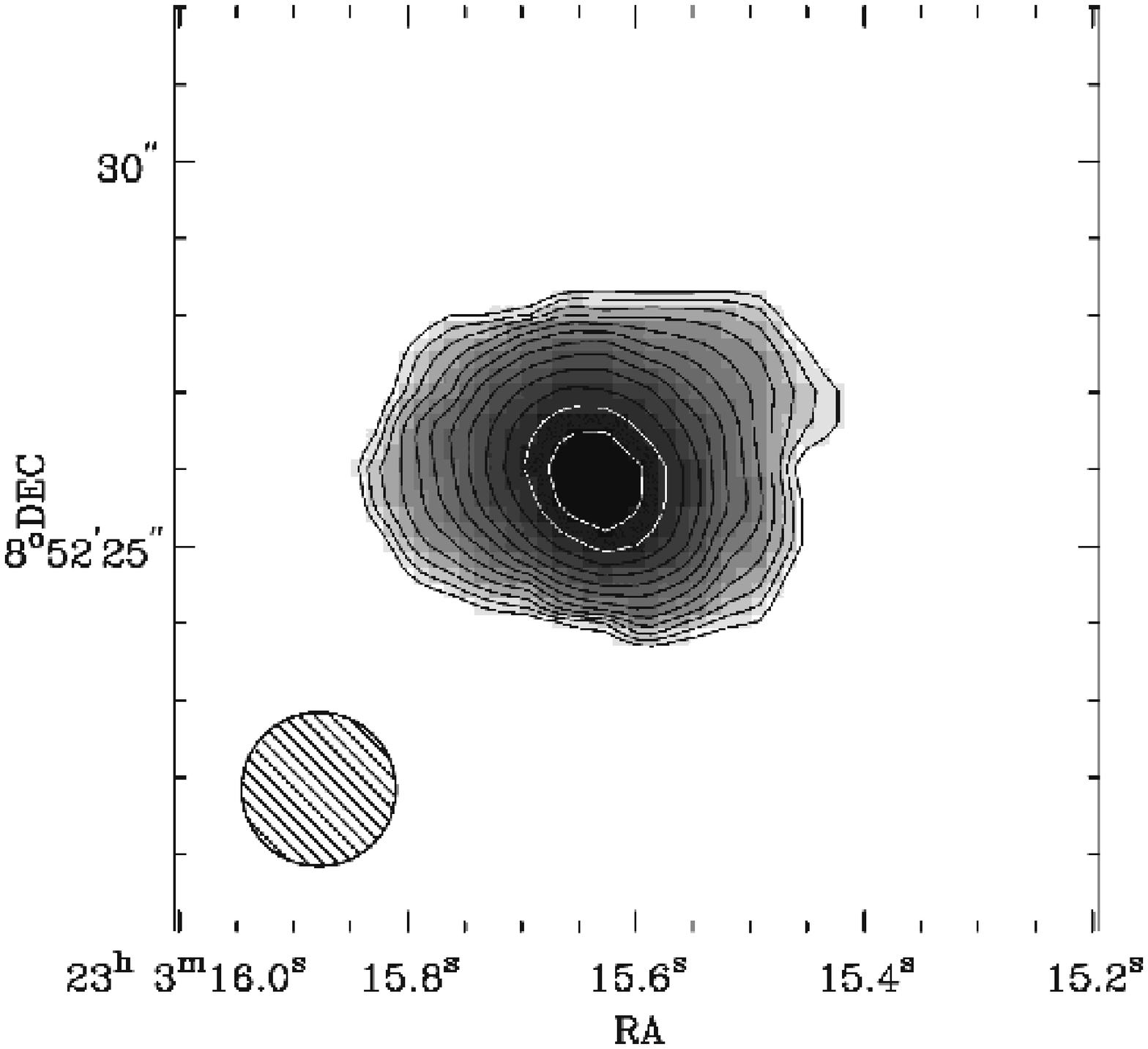,width=8cm}}
\caption{Integrated radio CO\,2-1 and HCN\,1-0 maps, at resolutions of
0.7\arcsec\ and 2.0\arcsec\ repsectively (beams shown to lower left
corners).
Contours and greyscales are logarithmic.
In the CO\,2-1 map, the greyscale and contours lie in the range
1-25.1\,Jy\kms.
The HCN\,1-0 map is drawn in the range 0.25--3.98\,Jy\kms.
}
\label{fig:COmap}
\end{figure}

The main features in the 0.7\arcsec\ resolution CO\,2-1 map in
Fig.~\ref{fig:COmap} are a very
bright marginally resolved central source, a partial ring or spiral arms 
at a radius of 1.5--2.5\arcsec\ (500--800\,pc), and a bright 
bar-like structure leading from the arms to the central source.
The regions of CO emission are in
general not coincident with the knots of K-band continuum emission,
which trace the active star forming areas.
This is easily seen in Fig.~\ref{fig:StarsGas}, which overlays the
stellar light in green on the CO emission in red.
We have used the {\em HST} J-band image for the
stellar light because the contrast of the star forming knots is
stronger than in the K-band, although in both bands the knots occur
in the same places.
For this figure, and in the analysis presented here, we have assumed
that the nuclear near infrared and CO peaks are aligned.
There is some suggestion that the dynamical centre and the CO peak may 
be offset by $\sim$0.1\arcsec\ (which may be expected if, as we show
in Section~\ref{sec:kin}, the nuclear CO is distributed in a ring) and
so it is possible that the CO and near infrared peaks are also
slightly misaligned.
However, because the offset is well within the resolution limits of
the CO data and we have no other way to perform accurate astrometry
for the spectroscopic data, we have ignored this small discrepancy.
The figure shows that along the major
axis (roughly north west) the star forming knots lie inside the ring
of CO emission, while along the minor axis they overlap.
The radio continuum images show a greater similarity to the CO map,
with bright star forming regions at
the ends of the bar (R2 and R3 in \citealt{col01}), and tracing the
same spiral/ring pattern as the CO\,2-1 out to $\sim2.5$\arcsec.

To compute the total line fluxes in the maps and the intensity-weighted 
mean velocity fields, we used uniformly weighted maps and ran the MOMENTS 
routine in the GIPSY package \citep{hul92}.
We did not include in the sum any emission below a limit of
2.5$\sigma$, and only included those features which were present in at
least 2 adjacent line channels.
The total CO\,1-0 flux we observed was 300\,Jy\kms, consistent with
that observed interferometrically by \citet{mei90}, and only sightly
less than the 400\,Jy\kms\ single dish flux of \citet{pap98}.
We observe a total flux of 470\,Jy\kms\ in the high resolution CO\,2-1 map.
\citet{pap98} measure a flux of $\sim$800\,Jy\kms\ with
both the NRAO 12-m and the James Clerk Maxwell telescopes, indicating
that we are missing 30--50\% of the single dish flux after accounting
for calibration uncertainties in all of the measurements.
This missing flux likely comes from extended faint emission which is
evident in the CO\,1-0 map \citep[see][]{tac96}.

The detection of HCN in the nuclear region of the galaxy indicates the
presence of dense gas, with molecular hydrogen volume densities of order
$10^5$\,cm$^{-3}$.
However, using the HCN/CO ratio to make quantitative statements is not
possible since the abundance ratio is unknown:
it typically lies in the range $10^{-4}$ in quiescent
clouds to $10^{-3}$ in star forming cores \citep{bla87}, while
\citet{ste94} have argued that in the nuclear regions of NGC\,1068 it
is more like $10^{-2}$.

\begin{figure}
\centerline{\psfig{file=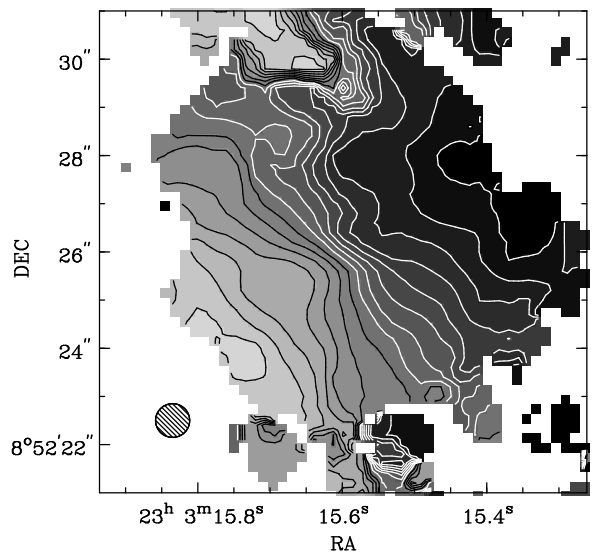,width=8cm,angle=-90}}
\caption{Velocity field for the inner regions of NGC\,7469, derived
for the CO\,2-1 line using the ROTCUR task in the GIPSY package.
Darker greyscales indicate negative velocities.
Contours are spaced at intervals of 20\kms.}
\label{fig:rotcur}
\end{figure}

We have quantified the motions in the central few arcseconds of
NGC\,7469 by two different methods.
First we solved for the velocity field using the ROTCUR task
\citep{beg89} in the GIPSY data reduction package.
This program divides the galaxy into a 
series of concentric tilted rings each having some circular velocity,
and makes least squares fits to the kinematic parameters of the galaxy
\citep{kru78}.
Although the program does not account for beam smearing effects, this
is only a concern in the inner arcsecond where the rotation curve is
steeply rising.
In regions outside this, beam smearing effects are not as serious.
We have fit for the rotation velocity and inclination angle 
by holding the kinematic position angle (PA=$128^\circ$), the systemic 
velocity ($V_{\rm LSR}=4925$\kms), and the position of the center of 
rotation constant as a function of radius.
We have not extended the fit beyond a radius
of about 3\arcsec.
The velocity field found by this task is shown in
Fig.~\ref{fig:rotcur}.

In such fits the inclination angle is an uncertain parameter
because it is degenerate with the actual rotation velocity.
We are unable to constrain it using the ROTCUR task, and instead
we consider the distribution of the CO emission knots around the nucleus.
If these form a pair of spiral arms, the pitch angle should have no
abrupt changes through the length of the arms \cite[e.g.][]{com95}.
Fig.~\ref{fig:COmap} clearly shows that in NGC\,7469, if the features
were continuous spiral arms, there would indeed be an abrupt change in
their pitch angle.
The alternative is to interpret the structure as a ring at a constant
radius of about 2.5\arcsec\ with a short pair of loosely wound spiral
arms or a bar leading to them from the nucleus. 
With such a configuration, the inclination must lie in the range
45--50$^\circ$, consistent 
with the 45$^\circ$ inclination of the outer stellar disk
\citep[e.g.][]{der86,mar94}. 
We therefore adopt an inclination of 45$^\circ$ in this paper.

We estimate the uncertainty in the position angle from the beam size
and the scale over which the fitting was performed, yielding
$\sim10^\circ$.
Inspection of the channel maps for the innermost CO data and also for
the HCN data, reveals that there is no evidence for a different
kinematic major axis between the ring and the nucleus:
within the uncertainties all the data are consistent with a single
major axis.
\citet{soi03} were able to resolve the nucleus of NGC\,7469 at
12.5\,$\mu$m using adaptive optics on the Keck telescope, and measured
a position angle of 135$^\circ$ for the emission on scales
of 0.1\arcsec.
This is consistent with the 128$^\circ$ we find, and supports our
conclusion that the ring on scales of 3\arcsec\ and the nucleus on
scales of 0.7\arcsec\ are oriented in the same plane.

\begin{figure}
\centerline{\psfig{file=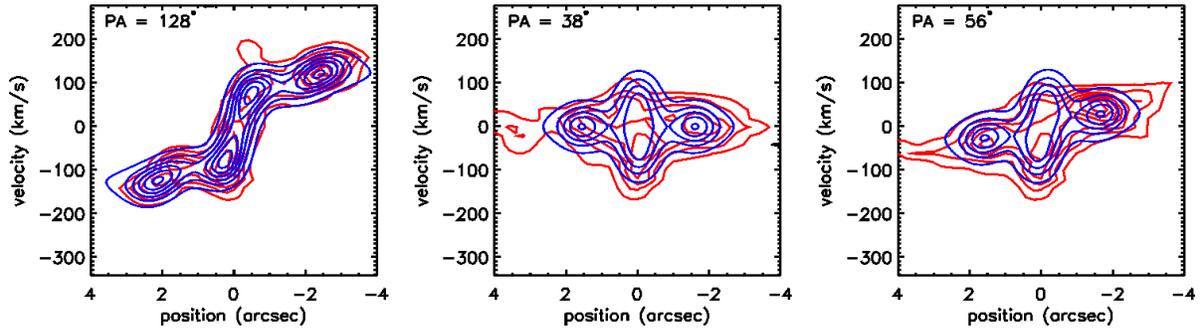,width=16cm}}
\caption{Position-Velocity diagrams for the CO emission at 3
position angles.
Left, 128$^\circ$ corresponds to the major axis;
center, 38$^\circ$ the minor axis;
and right, 56$^\circ$ lies along the apparent bar.
The observed data are drawn in red.
The axisymmetric model superimposed in blue shows very good agreement
with the data.
Note that there are none of the typical kinematic characteristics 
associated with a barred potential.}
\label{fig:COpv}
\end{figure}

The bar or spiral arm structure at a position angle of 56$^\circ$ is
unique to the molecular gas and
is evident neither in near infrared K-band images
\citep{gen95,lai99,sco00}, nor in sub-arcsecond resolution 8\,GHz
radio continuum maps \citep{con91,col01}.
If it were a bar, 
this would represent a very unusual configuration since stellar bars 
usually are
seen together with, and more prominently than, molecular bars
\citep[e.g.][]{reg99b,she02}.
It is, however, possible that the bar is an illusion, arising because
the spatial scales are small -- not much more than a few times the
spatial resolution -- which might lead to blending among different
clumps of CO emission.
This would be particularly noticeable along the minor axis, due to the
45$^\circ$ inclination, which lies close to the axis where the bar is
seen.
Such a possibility appears to be supported by
the fact that our data for NGC\,7469
exhibit none of the kinematic 
characteristics expected for a barred potential.
The velocity field in Fig.~\ref{fig:rotcur} shows no evidence for the
$\cal S$-shaped contours that arise in such cases 
\cite[see][and references therein]{kru78};
on the contrary it is remarkably uniform.
Additionally, there is no evidence for a tilted-$\cal X$ pattern in
the position-velocity (p-v) diagram taken along the bar axis (right
panel in Fig.~\ref{fig:COpv}).
These are characteristic of bar driven dynamics, and arise from the
high velocities associated with the $x_1$ and $x_2$ orbits parallel
and perpendicular to the bar.
Because the orientation of the bar is less than 20$^\circ$ from the
minor axis, the high speed motions in the $x_2$ orbits will be
correspondingly close to the plane of the sky, and so may not be
readily apparent in a p-v diagram;
on the other hand, the $x_1$ orbits should be more easily detectable.
However, the same blending as above, which might result in the illusion of
a bar could also conspire to make non-circular motions extremely
difficult to detect.
Furthermore, the relatively bright disk component and the large
turbulent velocity (see Section~\ref{sec:kin}) of 60\kms\ FWHM would tend
to swamp and hide the effects of non-circular motions.
An alternative might be that the structure is a pair of loosely wound
spiral arms, which would be associated with smaller non-circular
motions \citep{teu02} that might not be detectable in our data.
Without a more complete dynamical analysis we can neither
rule out nor confirm 
whether this apparent structure might be a bar or pair of spiral arms.
We therefore defer a discussion of non-circular motions
associated with spiral arms or a bar until later, and 
assume initially in our dynamical analysis in Section~\ref{sec:kin}
that the kinematic structure can be modelled axisymmetrically.

\subsection{Distribution and Properties of the 1-0\,S(1) Emission}
\label{sec:h2line}

As has been seen in Fig.~\ref{fig:2dspec}, 
strong narrow 1-0\,S(1) emission was observed across the nucleus of
NGC\,7469 at both position angles.
We have measured the flux and velocity at each point along the slit by
fitting a Gaussian line profile to the 30 pixels (1800\kms)
around the line in each row of the spectrum.
Initially, the velocity dispersion was left as a free parameter.
However, in every row of both position angles the derived value was
consistent with a constant 210\kms\ FWHM.
We therefore fixed the dispersion at this value.
Correcting for the instrumental resolution yields an intrinsic
dispersion of 185\kms\ FWHM.

The relative flux and velocity are shown in Fig.~\ref{fig:H2pv}.
The velocity zero point was taken as the velocity measured in the row
where the continuum was a maximum, and was 4965 and 4950\kms\ for the
2 position angles.
The uncertainty in each velocity measurement was determined by
constructing a 30 pixel segment, which had the same noise as the
original residual spectrum (after subtracting the Gaussian), and
to which the fitted Gaussian was added.
A new Gaussian profile was then fitted to the segment, producing
slightly different parameters.
The procedure was repeated 100 times, providing an estimate of the
uncertainties, which are shown as the error bars in the figure.

The total flux in a 0.5\arcsec\ length of the slit is
$2.4\times10^{-19}$\,W\,m$^{-2}$ and $1.6\times10^{-19}$\,W\,m$^{-2}$
for PAs 33$^\circ$ and 100$^\circ$ respectively.
The H$_2$ emission is easily resolved with FWHM of 0.45\arcsec\ and
0.33\arcsec\ at PAs 33$^\circ$ and 100$^\circ$ respectively.
These sizes are several times larger than the spatial resolution, and
so the difference can not be due to changes in the resolution on the
two nights (see Section~\ref{sec:contslope}).
We conclude that, from our line of sight, the H$_2$ emitting region is
asymmetrical, although with data from only two slit position angles,
we cannot fully constrain its size and shape.
An H$_2$ 1-0\,S(1) image of NGC\,7469 using the adaptive optics
system on the Canada-France-Hawaii Telescope was presented by
\cite{lai99}.
However, the difficulties of continuum subtraction, differing
resolutions, and the bright nucleus, mean that no morphological
information is available in the inner 0.5\arcsec.

\subsection{A Single Kinematic Model for CO and H$_2$}
\label{sec:kin}

Following the method outlined in \cite{tac94}, we have 
constructed kinematic models which can explain the various features
observed in the kinematic major and minor axis  
p-v maps, and which also estimate the rotation velocities
inside a radius of 5\arcsec.
As a starting point, these models use the position angle and
inclination derived previously and assume that all the gas lies in a
thin disk in the plane of the galaxy.
We then work towards an axisymmetric mass
surface density model which matches the observed velocities in the p-v
maps as closely as possible.
Because the process converges to the final solution iteratively, the
initial radial distribution is not critical.
Ideally, to reduce the number of iterations one would consider the
full mass distribution: of the stars (young and old populations) and
the gas.
However, this requires that the gas mass (i.e. the CO-to-H$_2$
conversion factor) is known {\em a priori} --
and we derive it afterwards in Section~\ref{sec:CO-H2}.
As a starting point, if one assumes that the conversion factor is
similar to 
that for the Galaxy, one finds that the gas mass dominates the total
mass within a radius of a few arcsec.
As we show in Section~\ref{sec:CO-H2}, in the central
regions the molecular gas does indeed account for more than half of
the total mass.
In particular, if one considers only the region where there are
recently formed stars, the annulus 1.3--2.3\arcsec, we find that the
mass of molecular gas is approximately 4 times that of the young
stars.
We therefore use an initial radial mass distribution for our model
based on the observed CO distribution.


The model allows us to define the mass surface density as a function of
radius using any number of Gaussian distributions, each
characterised by a relative scaling, a FWHM, and a radial offset.
The mass of the whole system is then scaled to that needed to match the
observed velocities for the given inclination.
In the last two steps, the distribution is convolved spatially and
spectrally.
The spatial kernal is identical to the beam (0.7\arcsec\ for the CO,
0.085\arcsec\ for the H$_2$).
The spectral kernal represents the turbulent velocity, that is random
motions on scales smaller than the beam.
If this is small, the kernal can instead be used to represent the effect of
the instrumental spectral resolution.
We compare the model p-v distributions with the data, and adjust the
radial mass distribution appropriately, iterating until there is
sufficient agreement between model and data.
This is a qualitative assessment, rather than a formal measurement of
chi-squared. 
The reason is that in this situation --- where the model is
necessarily much simpler than the data --- a chi-square value is often
strongly affected by inhomogeneities in the data (which the model is
not expected to duplicate) and a minimisation
routine can easily converge to a model that a person would clearly see is 
wrong.
We have therefore iterated on the model until it matches the main
features in the data as well as possible.
Using this method we have derived a single mass model
which matches 
the kinematics from both the CO radio data and the near infrared H$_2$
spectroscopic data at their different resolutions.

\begin{figure}
\centerline{\psfig{file=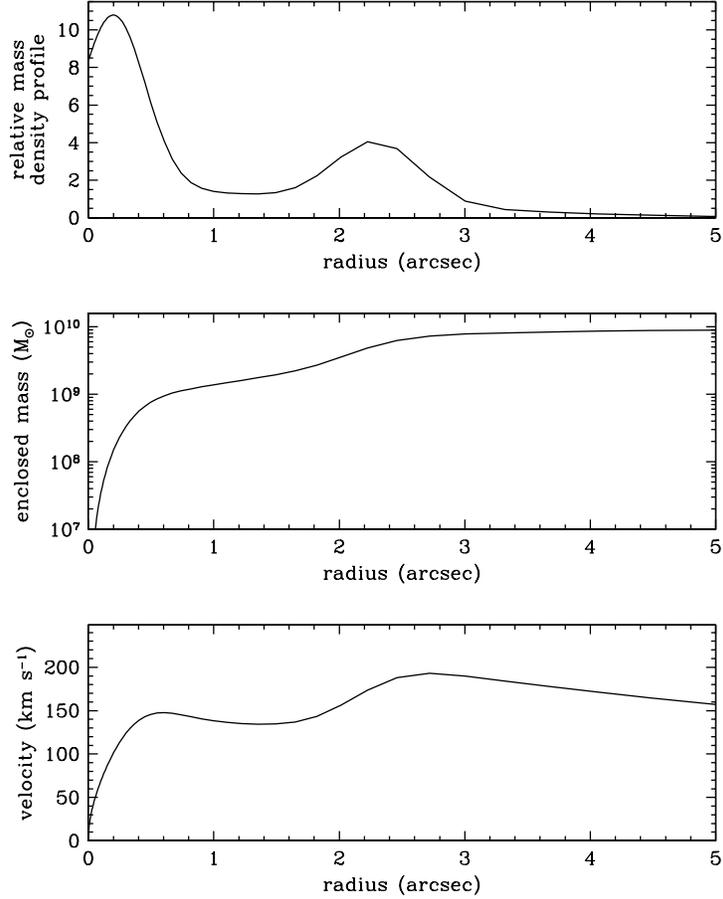,width=10cm}}
\caption{1-dimensional plots of the azimuthally symmetric mass model
comprising the 3 
components in Table~\ref{tab:massmod}, constructed to match both the radio CO
and the near infrared 1-0\,S(1) data of NGC\,7469.
At this stage, the profiles have not been convolved with a spatial beam.
Upper: mass surface density profile; the
extended nucleus, disk, and ring components are clearly distinguishable.
Centre: the resulting enclosed mass as a function of radius; 
the total mass is scaled to $9\times10^9$\,M$_\odot$.
Lower: rotation velocity as a function of radius in the plane of the galaxy.}
\label{fig:massmod}
\end{figure}

We have found that 3 components are needed to adequately match
the data, scaled to a total dynamical 
mass of $9\times10^9$\,M$_\odot$ within a radius of 5\arcsec.
These are summarised in Table~\ref{tab:massmod}, and also in
Fig.~\ref{fig:massmod} which shows the mass surface density profile, the
enclosed mass, and the resultant velocity in the plane of the galaxy
as functions of radius. 
The broad disk component is represented by a Gaussian centered on the
nucleus at $R=0$\arcsec\ with a FWHM of 4.8\arcsec;
the ring is defined by a Gaussian at $R=2.3$\arcsec\ that has a
FWHM of 0.8\arcsec, and with a peak intensity approximately twice that
of the disk component;
the nucleus itself is represented by a Gaussian at $R=0.2$\arcsec, a
FWHM of 0.6\arcsec, and a peak approximately 4 times the disk
component.
Additionally, a turbulent velocity of 60\kms\ FWHM for the CO, and
185\kms\ FWHM for the 1-0\,S(1), was needed to match the observed line
widths.
That the linewidth is much higher for the 1-0\,S(1) may be because 
the near infrared H$_2$ emission traces a hotter and more turbulent
component of the molecular gas.
The p-v diagrams for the major and minor axis of the CO emission are
shown in Fig.~\ref{fig:COpv}.
The equivalent diagrams for the H$_2$ emission are in
Fig.~\ref{fig:H2pv}.
In the latter figure, the model is drawn as a single line rather than
contours because otherwise the large line width makes it difficult to
compare to the data.
The velocity dispersion is dominated by the turbulent component
and hence is nearly constant across the central regions, as measured
in the data.

The nuclear component of the model can be interpreted in 2 ways:
either there is an
approximately uniform mass distribution within a radius of about
0.3\arcsec\ that then falls off like a Gaussian; or there is an
inner ring peaking at a radius of about 0.2\arcsec\ -- with the caveat
that if the innermost H$_2$ lies in a ring, its rotation curve yields
a measure of the total mass inside that radius but provides no
information about its distribution there.
Although the dynamical model cannot distinguish between these two
options,
we show in Section~\ref{sec:starclus} that most of the mass inside
this radius is due to stars rather than gas -- and hence that the gas
itself does lie in a ring.
Additionally, the interpretation as a ring is more readily understood
with respect to what is known about other similar galaxies.
For example, CO observations of NGC\,1068 \citep{hel97,tac97,sch00},
which is much 
closer at a distance of only 14.4\,Mpc, indicate that
this galaxy has a circumnuclear ring or spiral arms at $\sim$1\,kpc,
a nuclear ring at 70--100\,pc, and a prominent bar (stellar and gas)
between them.
Our dynamical model of NGC\,7469 suggests there is a circumnuclear
ring at $\sim$750\,pc and a nuclear ring at $\sim$65\,pc, perhaps with
short spiral arms or a bar between them.

\begin{figure}
\centerline{\psfig{file=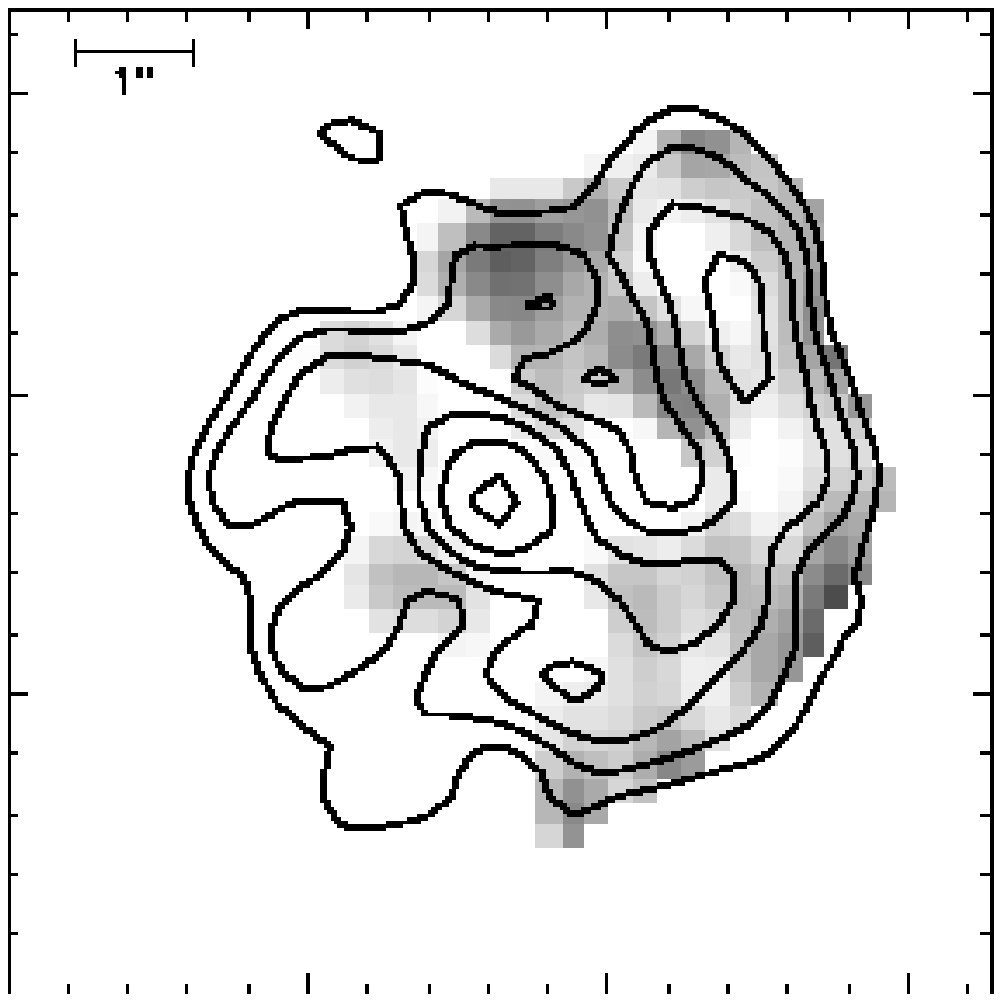,width=7cm}\hspace{1cm}\psfig{file=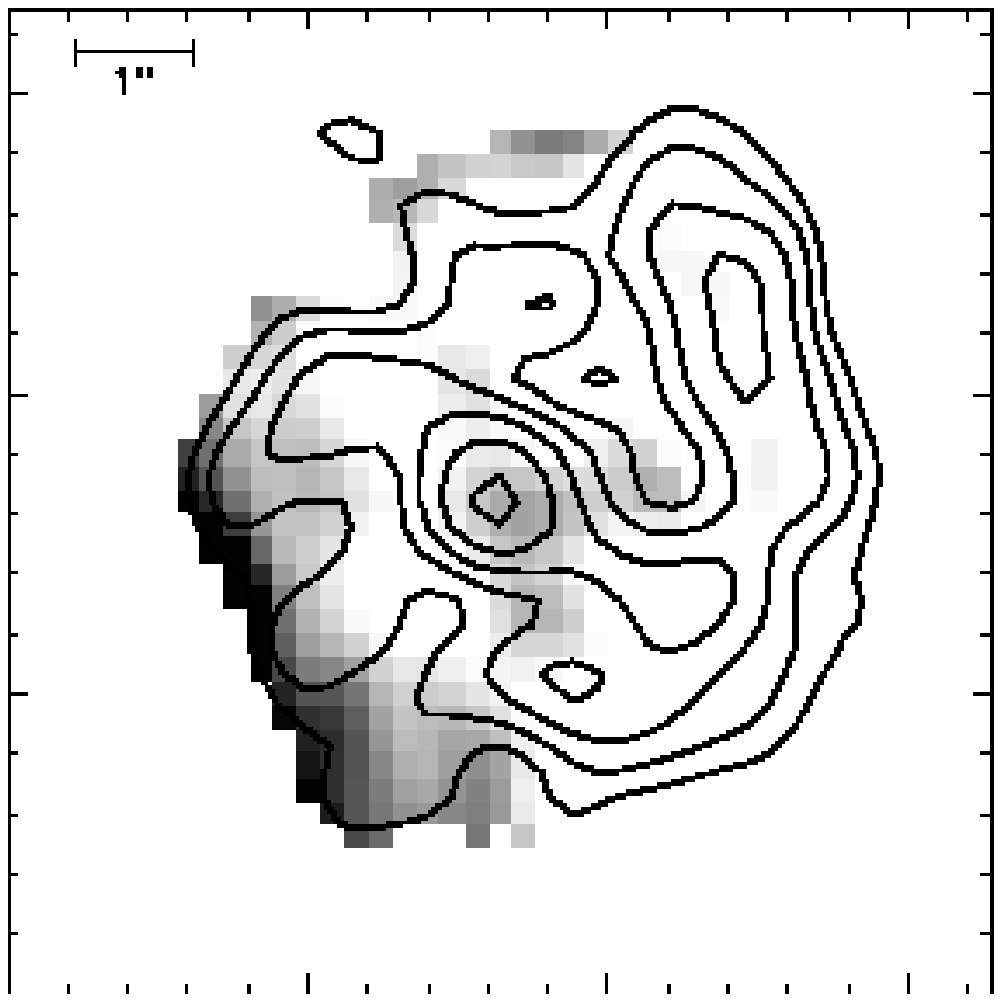,width=7cm}}
\caption{Residuals in the CO\,2-1 velocity field after subtracting the
axisymmetric model from the data. Positive residuals are in the left
panel, negative residuals in the right panel;
both are scaled from zero to 3$\sigma$ (75\kms).
The contours trace the flux distribution to indicate where the
molecular ring and possible bar or spiral arms lie.
The strongest residuals lie at the edges of the map where the flux is
weakest and hence the uncertainty greater.
There are no structured residuals above the noise limit along the bar
axis.}
\label{fig:res}
\end{figure}

In NGC\,1068, \citet{hel97} finds
non-circular motions associated with the bar and the spiral arms.
Along the arms, the typical streaming motions show an ordered
structure but are less than 30\kms;
in the region of the bar the kinematics are dominated by
non-circular motions.
For our CO data on NGC\,7469, a map of the velocity residuals, after
subtracting the model from the data, is shown in Fig.~\ref{fig:res}.
The standard deviation is 25\kms, and the strongest residuals are seen
around the edges of the map where the flux is weakest and hence the
uncertainties greater.
There is no obvious residual structure along the bar axis, which is
perhaps surprising given that, for example, \cite{roz00} see
a residual gradient from $+40$\kms\ to $-30$\kms\ along the bar in the
ionised gas of NGC\,3359.
It would appear therefore that we can rule out the presence of a
strong bar, which can result in residuals of at least this magnitude.
On the other hand, spiral arms, which give rise to residuals of only
10--50\kms, may be present \citep{teu02}.
We assume that the bar or spiral pattern implied by the morphology
is a real feature, but that the associated non-circular motions are
too weak for us to detect and that there is little gas inflow in the
region. 
Indeed, according to the conventional understanding, spiral arms or a bar are
expected to exist if the rings are interpreted as occuring at specific
resonances associated with spiral density wave theory.
Under the assumption that the deviations from axisymmetrical orbits
are small, we can use standard theory to estimate the pattern speed
of the bar/spiral.

The epicyclic approximation \citep[e.g.][]{bin87} asserts
that stellar orbits can be represented by the superposition of a
prograde circular orbit with 
angular velocity $\Omega(r)$ and a retrograde elliptical orbit with
frequency $\kappa$ where
$\kappa^2 = r{\rm d}\Omega^2/{\rm d}r+4\Omega^2$.
In an inertial frame the motion of these orbits forms unclosed
rosettes.
If a nuclear ring forms, it will do so in the vicinity of the inner
Lindlad Resonances (ILRs) which occur where, in the frame
rotating with the spiral pattern or bar with angular velocity
$\Omega_p$, the orbits close up on themselves.
Thus ILRs occur where 
$\Omega_p = \Omega - \kappa/2$.
Co-rotation (CR) occurs where the pattern speed is the same as the
orbital speed, $\Omega_p = \Omega(r)$.
Recently, \cite{reg03} have shown that more fundamental than the ILRs
to the existence of a nuclear ring is in fact the presence of
$x_2$ orbits (perpendicular to the bar).
A ring will form if the phase space filled by $x_2$ orbits is
sufficiently high that the gas on these orbits interrupts the flow along
$x_1$ orbits (parallel to the bar) where the two families of orbits
intersect.
The location of the nuclear ring moves inwards over
time, as a result of inflow of gas from further out along the bar; 
and hence the location of the ring cannot necessarily be used to
constrain the rotation curve of the galaxy.
However, we have shown above that in the circumnuclear region of
NGC\,7469 the non-circular motions are weak and that there is little
gas inflow.
For a weak bar, the innermost and outermost extent of the $x_2$ orbits
along the bar's major axis are well represented by the location of the
ILRs.
On the other hand, \citeauthor{reg03} argue that for strong bars the
concept of ILRs is not really applicable 
since the $x_1$ and $x_2$ orbits form closed loops in the bar's
reference frame along their entire extent.
How their theory relating the nuclear ring to $x_2$ orbits rather than
ILRs  should be applied to the case of multiple bars is not
clear.
\cite{mac00} showed that a secondary (inner) bar results from
a decoupling of the inner $x_2$ orbits:
the $x_1$ and outer $x_2$ orbits behave as expected for the primary bar,
while the inner $x_2$ orbits now form the backbone of the inner bar.
That is, they are equivalent to its $x_1$ orbits.
But it is not immediately obvious which orbits could then act as
the equivalent of the inner bar's $x_2$ orbits, thus allowing a
secondary nuclear ring to form.
Naively one would conclude that because there is no equivalent of
$x_2$ orbits for the inner bar, then there can be no nuclear ring
associated with it.
However, this must happen since there are many observations of
nested rings and bars.
It is certainly clear that our understanding of bars, rings, and gas
inflow is still developing.
Nevertheless, under the assumption that the non-exisymmetric
perturbations are relatively small and that the concept of ILRs
retains some meaning, it is helpful to consider the rotation
curve of NGC\,7469 in the context of the epicyclic approximation.

\begin{figure}
\centerline{\psfig{file=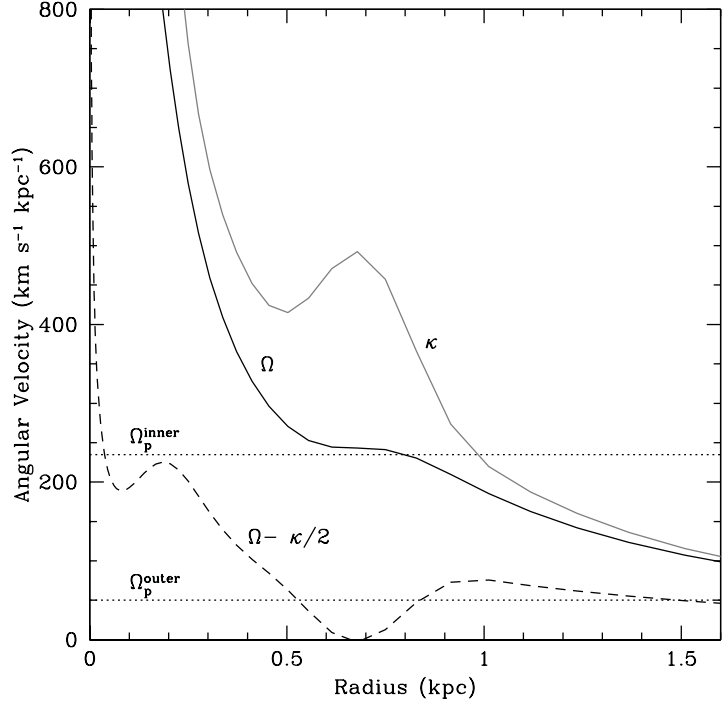,width=10cm}}
\caption{Rotation curve $\Omega(r)$ of the inner 5\arcsec\ of the
model shown in Fig.~\ref{fig:massmod}.
Also shown are the epicyclic frequency $\kappa$ (grey curve) and 
$\Omega-\kappa/2$ (dashed curve).
If the circumnuclear ring  at 2.5\arcsec\ (800\,pc) occurs at an ILR
of the large scale bar, and 
its position coincides with CR of the inner pattern speed
($\Omega_p^{inner}$), then to
also have a ILR at about 0.2\arcsec\ (65\,pc) which could account for
the nuclear ring means that
$\Omega_p^{inner} \sim 230$--240\kms\,kpc$^{-1}$.
The value shown here for the pattern speed of the outer primary bar,
$\Omega_p^{outer}$, is representative and could be anywhere
$\lesssim80$\kms\,kpc$^{-1}$.
See the text for a discussion on the applicability of the ILR concept.
}
\label{fig:ILR}
\end{figure}

Fig.~\ref{fig:ILR} shows $\Omega(r)$ derived from
our model, together with $\kappa$ and $\Omega - \kappa/2$.
We emphasize that we are not trying to use the rotation curve to make
predictions about the presence of nuclear rings, or vice-versa;
instead we are using the rotation curve together with the observations
of nuclear rings to try 
and make quantitative statements about the pattern speed of the
spiral/bar structure.
From the figure we estimate that the pattern speed at radii of
0.5--2\arcsec\ is about 230--240\kms\,kpc$^{-1}$.
\cite{wil91} have argued that the circumnuclear ring at
2--3\arcsec\ arises at the ILR of the large
scale bar.
If this is the case, then the ILR of the outer pattern occurs in the
vicinity of the CR of the inner pattern, and we expect that the
pattern speed of the primary bar is 
likely to be less than $\sim$80\kms\,kpc$^{-1}$.
In the standard model of \citet{mac00}, such a spatial coincidence is
required in order to reduce the number of chaotic zones around the
resonances, and may in fact be natural in such dynamical systems
\cite[e.g.][]{tag87}.
However, \cite{mac02a} argues that such resonant coupling hinders gas
flow to the nucleus because there are no straight shocks along the
inner bar.
This could explain why we see no obvious streaming motions within the
central few arcsec of NGC\,7469.

A pattern speed of 200--250\kms\,kpc$^{-1}$ for the spiral/bar structure
in the inner 800\,pc may provide a natural explanation for why the
star forming knots seen in the near infrared images are located at
smaller radii than the ring of CO emission.
If the K-band image of \cite{sco00} is deprojected (using position
angle 128$^\circ$ and inclincation 45$^\circ$), the knots of star
formation lie between radii 1.3--2.3\arcsec.
We suggest that they lie
downstream of the spiral/bar pattern seen in the gas.
Along the pattern, the gas density is locally increased and hence star
formation is triggered.
If we assume that we are seeing the star clusters at an age of
10\,Myr (when they appear brightest in the K-band due to the
appearance of late type supergiant stars), the
spiral/bar pattern which triggered the star formation will have moved
with respect to the stars in their orbits.
Taking the angular velocities for the orbits and pattern from
Fig.~\ref{fig:ILR}, we estimate that
the distance amounts to 0.4--0.5\,kpc (about 1/5 of a full circle) at
a radius of 0.4\,kpc (1.3\arcsec), 
but less than about 0.15\,kpc at a radius of 0.7\,kpc (2.3\arcsec)
since this is close to co-rotation.
Given also that the ages of the star clusters may differ from only a 
few\,Myr to more than 10\,Myr, one might
expect that the star forming knots are distributed a large part of
the way around the central regions of the galaxy, extending away from
the spiral/bar pattern but lying within the same radial range
(i.e. inside the ring of CO emission).
This is what the K-band images indicate.

Lastly, we briefly consider the black hole in the nucleus of
NGC\,7469.
It's mass has been 
estimated to be 3--$8\times10^6$\,M$_\odot$ by using the timescale of
the delay between UV and broad Balmer line variations together with the FWHM
of the broad lines \citep{col98}.
Given that our mass model already has $10^7$\,M$_\odot$ within a
radius of only 0.05\arcsec, 
the rotation curve is insensitive to the effects of such a low mass
black hole at our spatial resolution and signal to noise;
we can only distinguish models that have black hole masses exceeding
$5\times10^7$\,M$_\odot$.

\subsection{Excitation of the 1-0\,S(1) Emission}
\label{sec:h2exc}

\begin{figure}
\centerline{\psfig{file=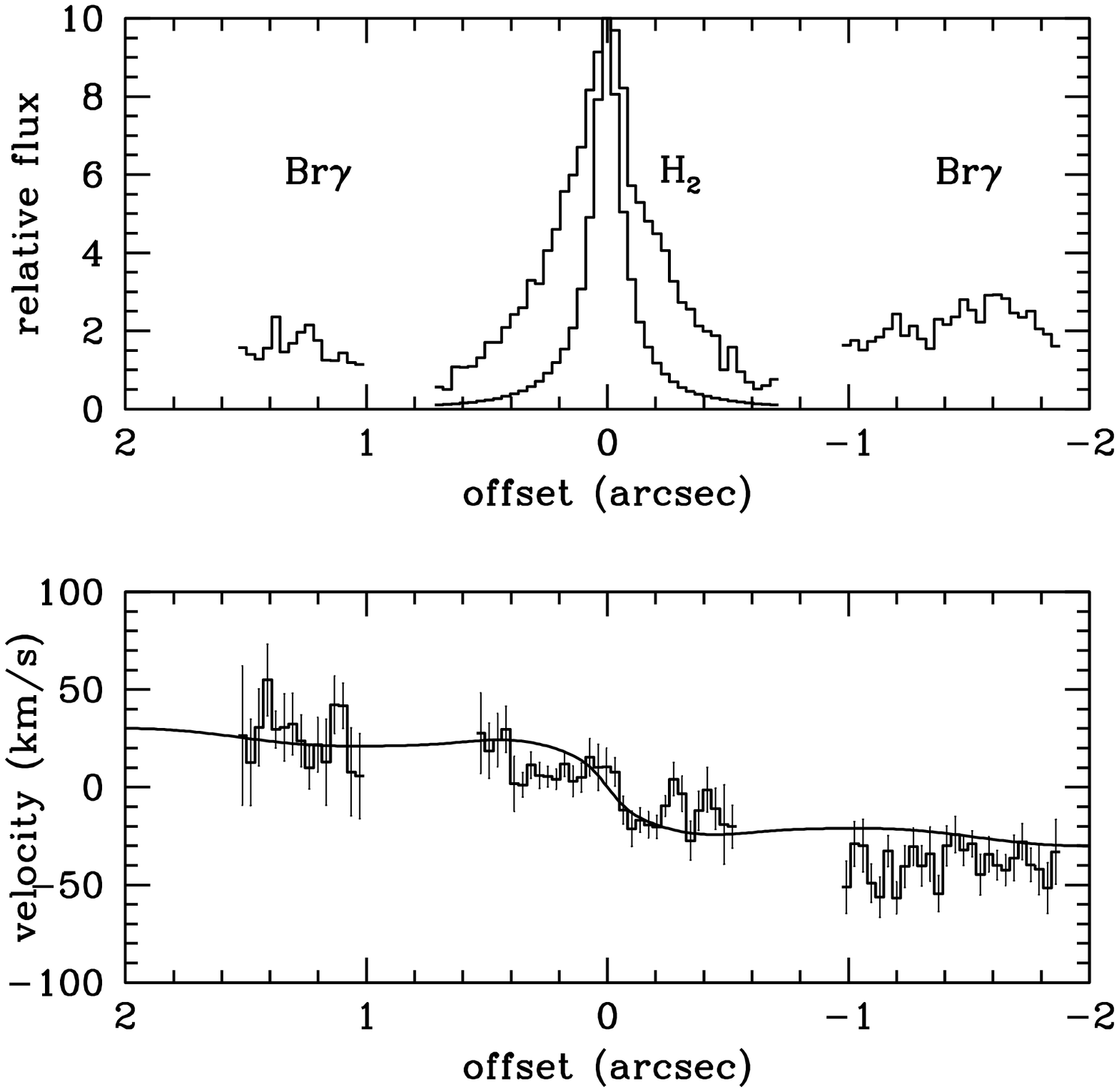,width=8cm}\hspace{5mm}\psfig{file=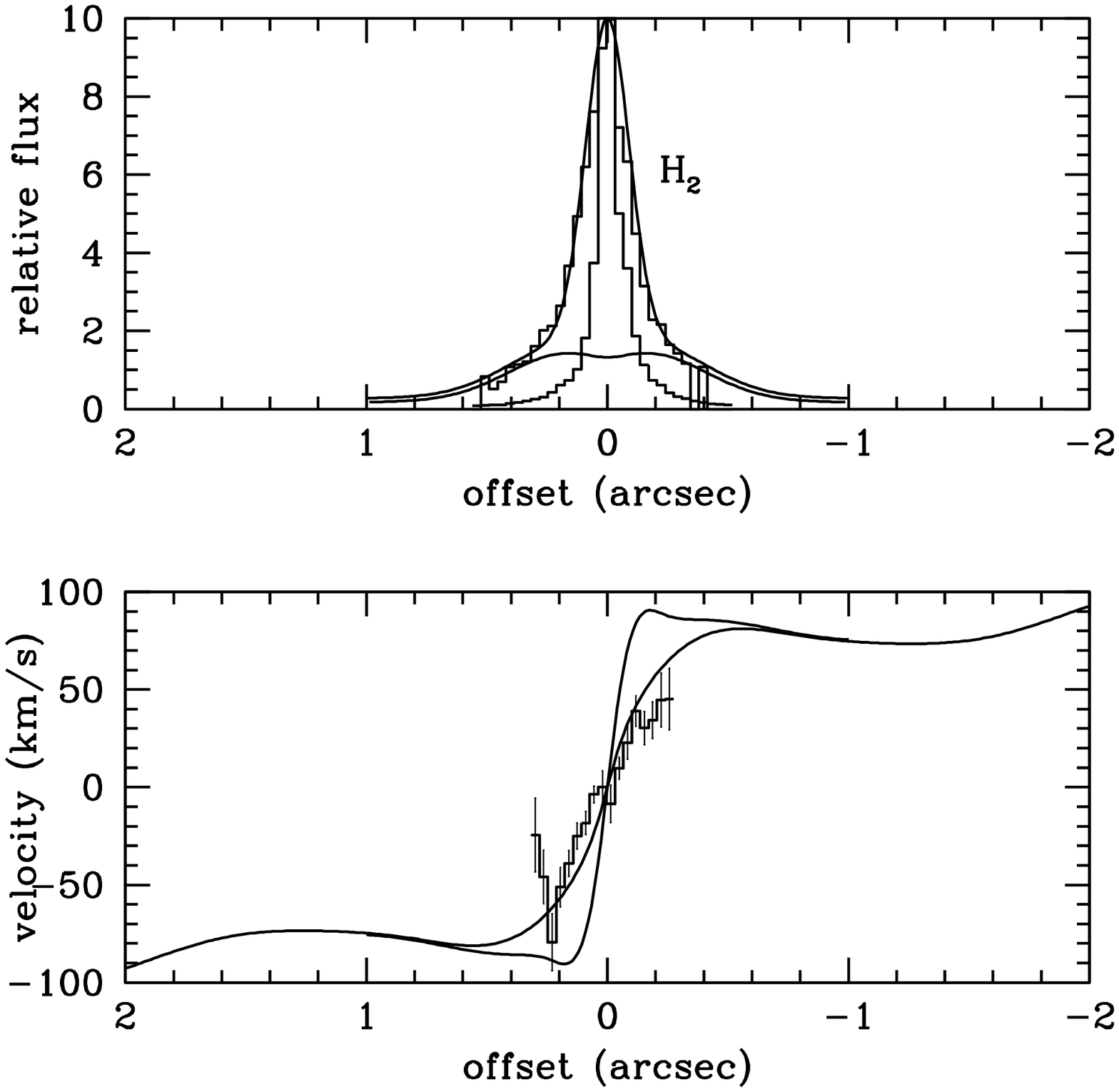,width=8cm}}
\caption{Flux (upper) and velocity (lower) profiles for
the H$_2$ and Br$\gamma$ lines at position angles 33$^\circ$ (left)
and 100$^\circ$ (right).
In the upper plots, the line flux profiles (H$_2$ or Br$\gamma$
as labelled) are drawn as solid histograms, while the continuum
profile (for comparison) is green.
The associated velocity measurements are also shown as solid
histograms, with 1\,$\sigma$ error bars derived as
described in the text.
The smooth red curve superimposed is the velocity profile of the
axisymmetric mass model.
Note that in the left hand panel, a model at position angle 20$^\circ$
was used instead of the 33$^\circ$ of the data; 
at the spatial scales involved, this is only a small discrepancy and
could be mostly due to the uncertainty in the estimate of the
kinematic major axis.
In the right hand panel, the 1-0\,S(1) flux profile expected from
the mass model (shown in red) differs dramatically from the observed
flux profile.
However, if an additional core component is included (blue flux
profile in upper right panel), and the masses scaled appropriately,
the derived velocity profile (blue curve in lower right panel) does
not match that observed.
The conclusion is that the 1-0\,S(1) lines does not trace the gas
distribution.
}
\label{fig:H2pv}
\end{figure}

It is remarkable that the radial profile of the 1-0\,S(1)
emission is very different from that of the mass distribution in the
model.
In Section~\ref{sec:kin} we showed that the velocity curves of the CO
and 1-0\,S(1) lines indicate that the mass distribution in the
nucleus is extended.
If the 1-0\,S(1) emission were to trace this distribution, the expected flux
profile would be much broader and flatter than that observed
(red curves in right panel of Fig.~\ref{fig:H2pv}).
We can rule out the possibility that the flux profile traces the
underlying mass distribution by considering a model in which this is
the case.
In such a model, in order to match the velocities at radii beyond
0.5\arcsec, the mass of the nuclear ring has to be reduced by one
third to 
$6\times10^8$\,M$_\odot$, and the mass lost from it replaced by a
nuclear core component of $3\times10^8$\,M$_\odot$.
The central part of the resulting rotation curve is much steeper
than that actually observed (blue curves in Fig.~\ref{fig:H2pv}).
Additionally, there would be a measurable increase in the velocity
dispersion from 185 to a maximum of nearly 210\kms\ FWHM over the
central 0.3\arcsec.
With our preferred mass model, the dispersion increases by no more
than a few \kms\ over the central 0.1\arcsec, consistent with the lack
of any variation seen in the data.
This result -- that in NGC\,7469 the 2 tracers of molecular hydrogen
CO and 1-0\,S(1) have different distributions -- is in contrast to
what is seen in NGC\,1068. 
In that galaxy both the CO \citep{sch00,bak00} and the 1-0\,S(1)
\citep{rot91,bli94,dav98} trace the same distribution: a circumnuclear
ring on scales of 70--100\,pc.

We suggest that at least some, and perhaps most, of the H$_2$ is
excited by X-ray and UV irradiation from the AGN.
By comparison to NGC\,1068, one would be surprised if this were not
the case:
in that galaxy, which has a similar bolometric luminosity, the AGN is
able to cause strong 1-0\,S(1) emission even in the circumnuclear ring
at a distance of 1\,kpc \citep{dav98}.
If the 1-0\,S(1) were excited in this way in NGC\,7469, then 
since the intensity of the X-ray and UV radiation would fall as $R^2$,
or even faster if dust 
attenuation is important, one would also expect a rapid decrease in the
1-0\,S(1) intensity, as observed.
Furthermore, the H$_2$ morphology would depend largely on
the details of the dust distribution and clumpiness, and would not
necessarily reflect the distribution of the cold gas.
One could test this conclusion further (Davies et al. in prep.) by
obtaining a deep spectrum of 
the H$_2$ lines to look for suppression of the 2-1\,S(3) line, which
\citet{kra00} claim to have seen in NGC\,1275 and which may be a
characteristic of X-ray irradiated gas \citep{black87,mal96}.


\subsection{CO-to-H$_2$ Conversion Factor in the Circumnuclear Region}
\label{sec:CO-H2}

The stellar mass within a radius of $2.5$\arcsec\ of the nucleus has
been estimated by \cite{gen95}.
In this Section, we estimate the dynamical and hence gas mass in the
same region in order to determine the CO-to-H$_2$ conversion factor
for the inner 1.6\,kpc of NGC\,7469.
We begin with the assumption that
\begin{equation}
M_{\rm dyn} \ = \ M_{\rm gas} \ + \ M_{\rm old} \ + \ M_{\rm burst}
\end{equation}
where M$_{\rm old}$ is the mass in the old stellar population and
M$_{\rm burst}$ is the stellar mass formed in a recent starburst. 

To derive the mass of old stars, \citealt{gen95} measured the J-band flux
density between the nucleus and the ring, and scaled this to the full
aperture.
Using a conversion factor between near infrared flux density and old
stellar mass, they estimated that $2\pm1\times10^9$\,M$_\odot$ is
in the form of old stars.
To determine the mass of recently formed stars, they used starburst
models to calculate the parameters of the starburst ring.
Their two favoured models imply young stellar masses of about
$7\pm3\times10^8$\,M$_\odot$.
Hence, the total stellar mass in this region is
$\sim2.7\times10^9$\,M$_\odot$.

In Section~\ref{sec:kin} we found that for an inclination of
45$^\circ$, the dynamical mass within 5\arcsec\ is
$9\times10^9$\,M$_\odot$;
within a radius of 2.5\arcsec\ it is $6.5\times10^{9}$\,M$_\odot$.
The difference between this estimate and that in \citet{gen95} is due
to their assumption of an isotropic velocity field for the late
type  stars causing the 2.3\,\micron\ CO absorption.
If instead the stars move primarily in a low inclination disk plane
(a logical conclusion for a young starburst in which the red
supergiants dominate the 2\,\micron\ emission), the observed velocities and 
velocity dispersions become smaller than the intrinsic ones and
the discrepancy vanishes.

The mass estimates above lead us to conclude that the total gas mass
in the 
same region is $\sim3.8\times10^9$\,M$_\odot$,
more than the total mass of stars and several times more than
the mass of recently formed stars in the ring.
It is therefore apparent that there is still sufficient gas to continue
forming stars for a considerable time.
To get to the molecular gas mass, we assume that the masses
of H{\small I} and H{\small II} are negligible contributions to the
gas mass at radii interior to 
2.5\arcsec, and correct for the mass due to helium (40\%).  
We find a H$_2$ mass of $2.7\times10^9$\,M$_{\odot}$ within
this region.

We can now use this mass to derive a CO-to-H$_2$ conversion factor
and compare it to 
the Galactic conversion factor between N(H$_2$) and CO integrated intensity 
of $X_G=2.3\times10^{20}$\,H$_2$\,cm$^{-2}$\,(K\kms)$^{-1}$ 
\citep{str88}.
In terms of the mass of H$_2$, the relationship can be written as
\begin{equation}
M({\rm H_2})\,{\rm [M_\odot]} \ = \ 
	9.0\times10^3 \  
	X / X_G \ 
	L_{\rm CO}\,{\rm [Jy\kms]} \ 
	D^2\,{\rm [Mpc]},
\end{equation} 
where $D$ is the distance to the source.  
Our previous observation of NGC\,7469 in the CO\,1-0 line (with a
spatial resolution of 2.5\arcsec) with the IRAM interferometer yielded
a total flux of 300\,Jy\kms, roughly half that seen in the single dish
spectrum of \citet{hec89}.
A crude correction for the possibly missing flux can be made by
assuming that the remaining flux is distributed in a smooth source
$\sim20$\arcsec\ across, the size of structure which would have just
been resolved out by the IRAM interferometer.
The ``surface brightness'' of such a structure would be
0.8\,Jy\kms\,arcsec$^{-2}$.
Within a radius of 2.5\arcsec\ the CO\,1-0 flux is 100\,Jy\kms.
Including the correction for the possibly missing flux raises this to
a maximum of 116\,Jy\kms.
Solving for $X$ in the mass equation yields $X/X_G=0.59$.
The uncertainty of this value is not trivial to estimate.
Accounting for the stellar masses and missing CO flux puts it at about
30\%.
Uncertainty in the inclination is also important: a 5$^\circ$ change
will alter the dynamical mass by 17\% and hence the gas mass by 29\%.
Combining these we estimate a total uncertainty of about 40\% on the
conversion factor we have calculated, so that it could lie in the
range $X/X_G=0.4$--0.8.

There is mounting evidence that the CO-to-H$_2$ conversion factor in
intense star forming environments is less than that for the Milky Way.
\cite{sol97} and \cite{dow98} found that for ultraluminous galaxies
the difference was about a factor five. 
They argued that in such cases, it is the warm diffuse phase of
the CO that dominates the emission, rather than the denser clouds from
which the conversion factor is measured in the Milky Way.
This effect was also observed for NGC\,7469 by \cite{pap00}.
While we do not find a discrepancy as large as theirs, within the 
uncertainties our result tends in the same direction.

\section{The Nuclear Star Cluster}
\label{sec:starclus}

\subsection{As Measured by the CO Absorption}
\label{sec:COflux}

The nuclear K-band light from NGC\,7469 is a combination of thermal
emission from hot dust associated with the AGN and stellar light,
which can be traced through the CO\,2-0 bandhead absorption longward
of 2.29\micron\ in the spectra of late type stars.
Although the hot dust emission can dramatically reduce the equivalent
width of the stellar absorption features, it does not change their
``flux'' (absolute absorption).
Hence, under the assumption that the stellar types dominating the CO
features are invariant throughout the nuclear cluster, the spatial
extent of the CO flux directly traces the spatial extent of the
stellar cluster.

\begin{figure}
\centerline{\psfig{file=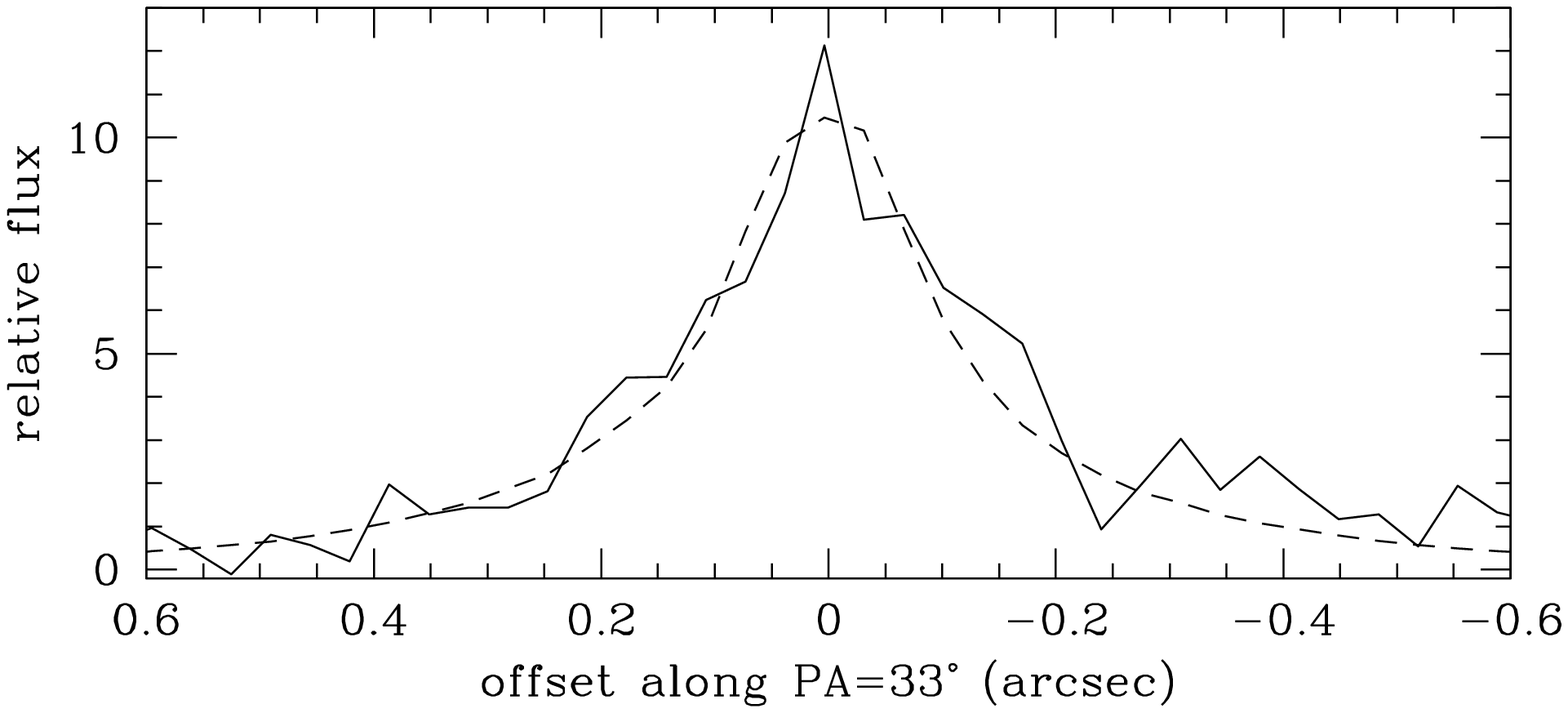,width=8cm}\hspace{5mm}\psfig{file=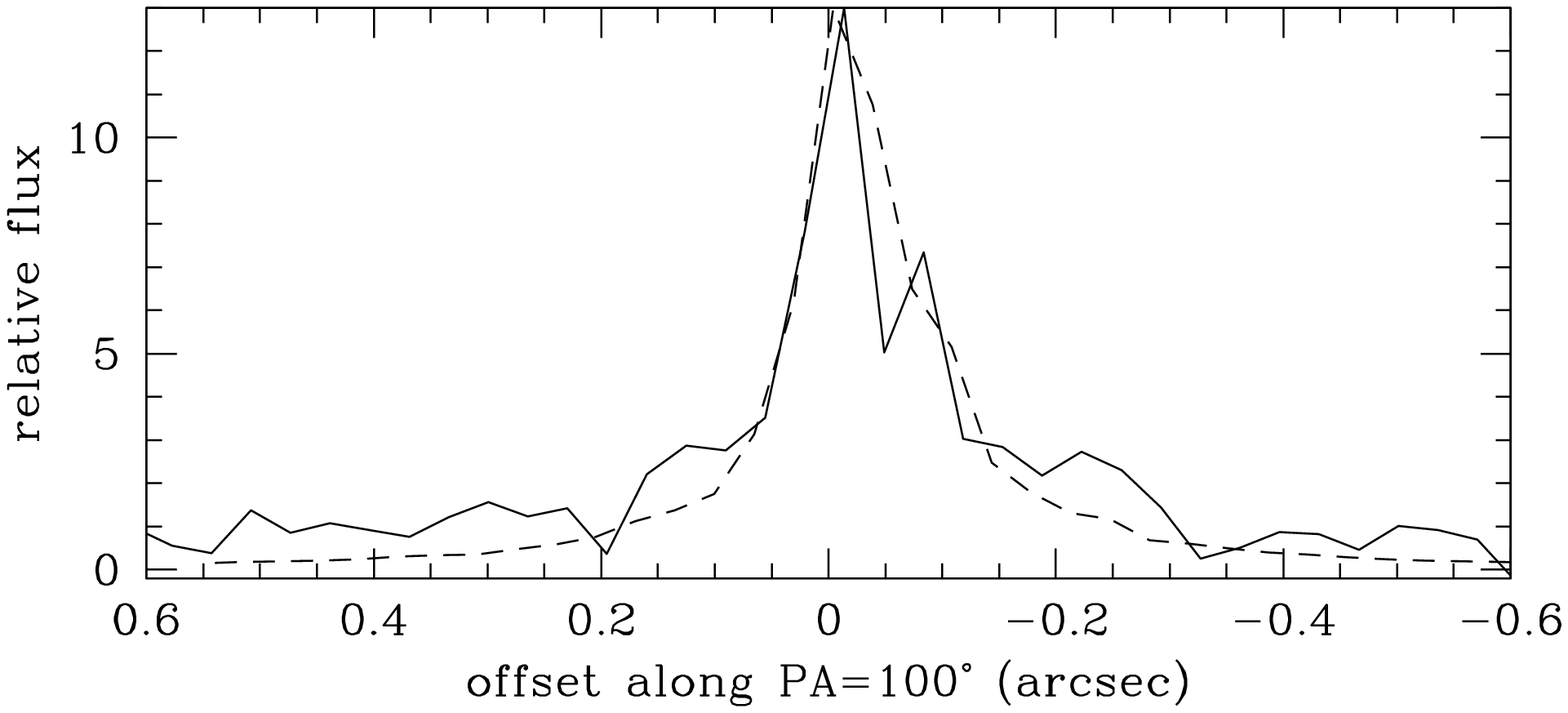,width=8cm}}
\centerline{\psfig{file=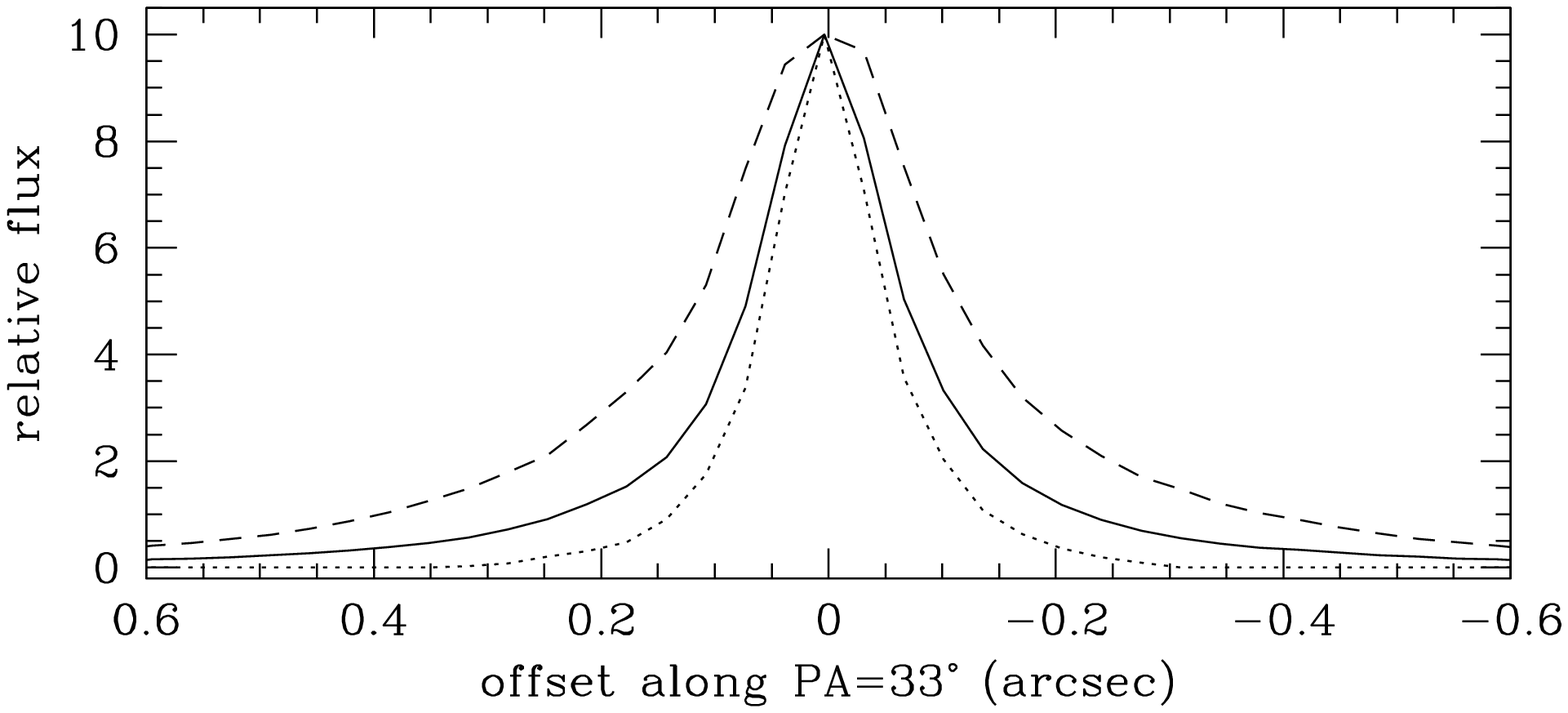,width=8cm}\hspace{5mm}\psfig{file=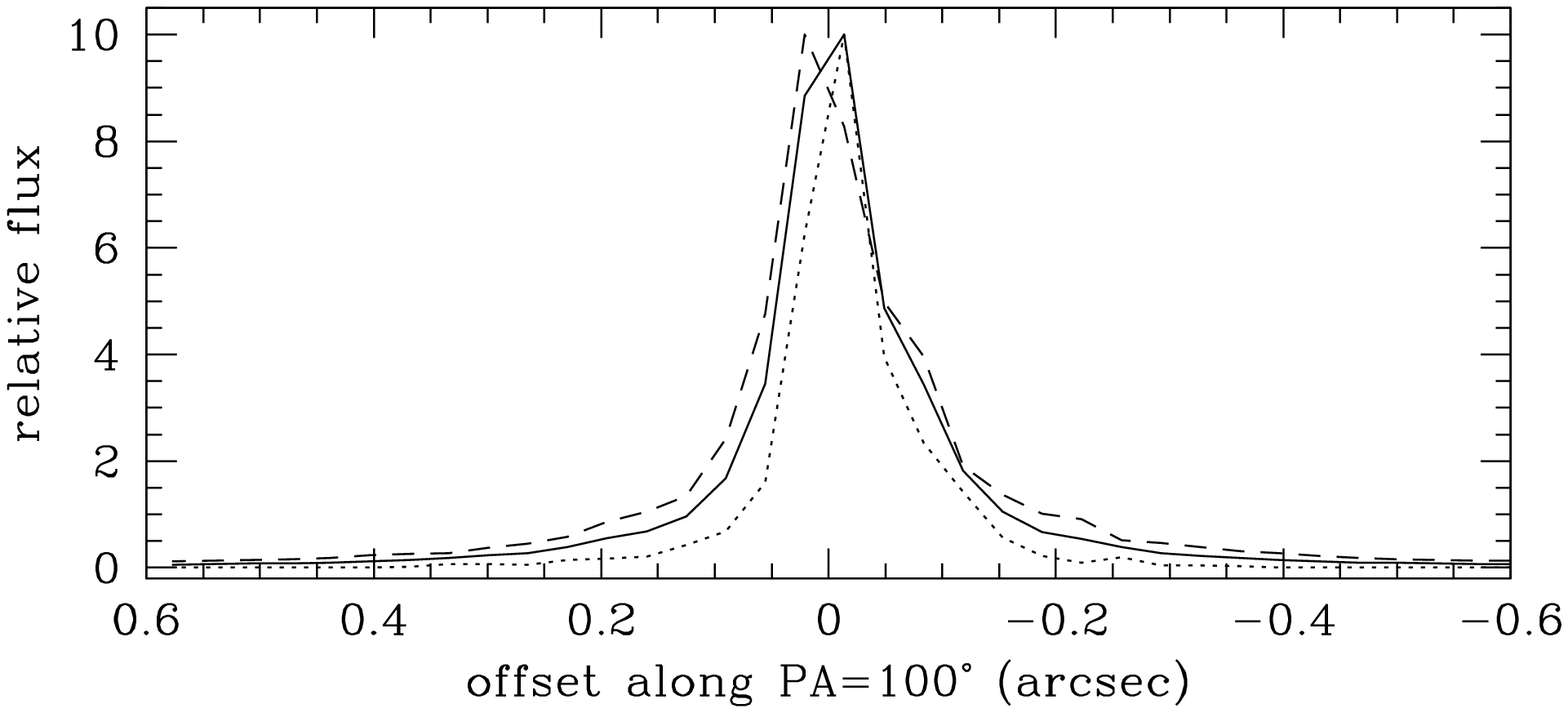,width=8cm}}
\caption{
Spatial profiles at position angles 33$^\circ$ (left) and 100$^\circ$
(right).
Upper: CO 2-0 bandhead flux (solid line), and stellar profile derived
from the continuum spectal shape (dashed line, see below).  
Lower: Profiles derived from fitting hot dust and a stellar template to
the continuum spectral shape.
The solid line is the observed continuum profile, the dashed line the
derived stellar cluster profile, and the dotted line the derived
profile of the hot dust core.}
\label{fig:starclus}
\end{figure}

To determine the extent of the CO absorption, we have measured the
flux in the wavelength range 2.293-2.302\,\micron.
The continuum level has been determined from an extrapolation of a
linear fit to the 2.086--2.293\,\micron\ range (excluding the region
around the Br$\gamma$ line).
The (relative) CO flux at both position angles is shown as the solid
line in the upper panel of Fig.~\ref{fig:starclus}, and has been
directly resolved in both cases.

The equivalent widths of the CO bandhead are similar at both
position angles, with
$W_{\rm CO} = 1.7$\,\AA\ in a 0.5\arcsec\ length of the slit.
By comparison to similar values for late type stars, one can in
principle estimate the fraction of stellar light in the nucleus.
However $W_{\rm CO}$ is very dependent on stellar type,
and hence on the age of any star forming episode.
To have a measurable CO absorption, a star cluster must be at least
10\,Myr old.
At this age late type supergiants dominate;
but as the age increases giant stars dominate more.
To quantify this effect, we have measured the equivalent width of the
CO bandhead in the same wavelength range for several different stellar
templates \citep[from][]{wal97} and find $W_{\rm CO}=5$--$30$\,\AA\
depending on stellar type, as shown in Table~\ref{tab:stars}:
$\sim5$\,\AA\ for late type dwarfs, $5$--$20$\,\AA\ for giants,
$20$--$30$\,\AA\ for supergiants.
These values suggest that stars could contribute anything in the range
5--40\% of the observed continuum at 2.3\,\micron.
Lower fractions would apply if supergiants dominate the
K-band continuum, and higher fractions for giants;
dwarf stars are unlikely to dominate the spectrum due to their
intrinsic faintness.


\subsection{As Measured by the Continuum Slope}
\label{sec:contslope}

An alternative and completely independent method to estimate the size
and brightness of the 
stellar cluster is to use the slope of the continuum.
We have adopted an iterative procedure to decompose the continuum into
stellar and thermal (hot dust) parts in each spatial row of the spectrum.
Initially, we assume that all the flux in the spatial row where the
continuum is a maximum is due to hot dust.
We can then make an initial estimate of its temperature by calculating
the spectral shape of the hot dust component and comparing it to that
observed. 
The second step needs in addition a stellar template of a late
type star. 
We use the K1.5\,Ib HR\,8465 from \cite{wal97}, although the exact
type does not matter as the slopes of all cool stars are similar.
We scale these two components to match as well as possible the
observed continuum spectral shape in each other row of the spectrum.
The stellar cluster is traced through the scaling applied to the
stellar template, and yields a smooth profile except for the few rows
near the centre.
One can interpolate over these to estimate what the stellar
contribution in this region should be.
Subtracting this from the continuum of the central row used in the
initial step leaves something that is
closer to the original assumption that (the remaining spectrum in)
this row is pure hot 
dust, and allows one to make a new estimate of the hot dust
temperature.
The procedure is repeated until the fit converges.
Because this process uses the whole slope of the continuum, it
provides a high signal-to-noise measurement of the spatial profiles of
the hot dust core and stellar cluster.
On the other hand, it is rather sensitive to errors in the slope of
either the object spectrum or the standard star used to calibrate out
atmospheric features.

Since extinction acts to modify the spectral slope, we need to
ask whether the results are affected by ignoring it.
For the stellar light, one can examine the spectral shape at an offset
of ~0.2\arcsec\ from the continuum peak -- where stellar light
dominates the emission but also where there is still enough flux for a
reliable measurement.
We find here that the measured continuum slope matches the stellar
template slope well, indicating that there is probably very little
reddening of the observed emission.
For the hot dust emission, there is a degeneracy between the
temperature of the dust and the extinction towards it.
Without extinction, we derive a characteristic temperature of
$T=600$\,K.
The spectral shape of this blackbody can be reproduced almost exactly
by reddening a blackbody with a higher characteristic temperature.
Thus, if sublimation limits the maximum temperature of dust grains to
$\sim1500$\,K, we can constrain the maximum foreground exinction
screen to have $A_{\rm V} \lesssim 35$\,mag.
Thus, while reddening can change the derived temperature of the hot
dust component, it does not affect the relative scaling between this
and the stellar light.

The results of the fitting procedure are shown in the lower panels of
Fig.~\ref{fig:starclus}.
The hot dust core (dotted line) has a narrower profile than the
observed continuum (solid line), and provides a measure of the spatial
resolution achieved.
At PA\,33$^\circ$ (first night), its FWHM is 118\,mas, compared with
146\,mas for the full continuum;
at PA\,100$^\circ$ (second night), its FWHM was 80\,mas (cf. 104\,mas
for the full continuum), similar to that measured on the PSF
calibrator.
The difference may be due to a better AO correction on the
second night.

The profile of the stellar cluster has been
plotted (suitably scaled) in the upper panels of the figure.
At both position angles, there is excellent agreement between this
profile and that measured from the CO bandhead; 
the measured FWHMs are
223\,mas at PA 33$^\circ$, and 115\,mas at PA 100$^\circ$.
This difference is much greater than that in the core noted above, and
cannot be 
due solely to the AO performance: at both position angles the stellar
cluster is clearly resolved.
Applying a quadrature correction for the finite resolution, one
derives an intrinsic FWHM of 0.189\arcsec\ at PA 33$^\circ$,
equivalent to 60\,pc at the distance of NGC\,7469.
Similarly, at PA 100$^\circ$ the size is 26\,pc.
These values are similar to the derived $\sim$45\,pc size of the stellar
cluster around NGC\,1068 \citep{tha97}.
However, unlike that cluster
the one in NGC\,7469 appears to be distinctly non-axisymmetric.
With measurements at only 2 position angles one cannot fully
determine the morphology, 
although based on the current observations the cluster does lie fully
within the region enclosed by the nuclear ring.

From the scaling of the stellar light component, we find that in a
0.5\arcsec\ length of the slit, it contributes $\sim40$\% of
the continuum at 2.3$\mu$m (at both position angles).
This is consistent with the upper end of the range derived from the CO 
flux in Section~\ref{sec:COflux}, as well as the work of \cite{maz94}
based on PAH emission.
These authors found that 25\% of the total 3.28\,$\mu$m
PAH emission occurs within a radius of 1\arcsec\ from the nucleus,
and they concluded that at least 1/3 of the nuclear K-band light in
this region originates in star formation.
A stellar light fraction close to 40\% would tend to suggest that, 
based on the equivalent widths measured in Section~\ref{sec:COflux}, 
it is late G and early K type giants which dominate the stellar light.
However, this is an implausible average stellar type.
Population synthesis indicates that once the supergiant phase is past,
an average type K4--5 III is more likely.
It is possible, therefore, that the ``continuum slope'' method above
has overestimated the stellar contribution, and that a fraction 
closer to 20\% is more realistic.

An alternative approach to estimate the stellar content of the nuclear
cluster is to use the stellar mass and K-band magnitude as constraints on
evolutionary synthesis models.
The mass model we derived in Section~\ref{sec:kin} indicates that 
within a radius of 0.10\arcsec\ (chosen to match the FWHM of the
nuclear star cluster) the total mass is only
$3.5\times10^7$\,M$_\odot$.
The mass of stars in the cluster cannot exceed this.
The  magnitude along the slit within $\pm0.25$\arcsec\ was found
in Section~\ref{sec:obs2} to be $K=14.0$\,mag; 
within $\pm0.10$\arcsec\ it is only 0.1\,mag fainter.
Extrapolating to a 0.2\arcsec\ circular aperture increases 
the brightness to $K=12.8$ (0.8\,mag fainter than for a 0.5\arcsec\
aperture).
The stellar light is effectively all included within such an aperture.
If we assume a Salpeter IMF, solar metallicity, and continuous star
formation, the
evolutionary synthesis model Starburst 99 \citep{lei99} allows us to
use the magnitude and mass limit above to
constrain the fraction of stellar light and the age of the star
cluster.
In order to avoid exceeding the mass, the K-band stellar light
fraction must be less than 50\%.
if it lies in a range 20--30\% the age of the star formation is
constrained to be 10--60\,Myr (older ages are only possible if the
stellar light fraction is lower).
This yields masses between $1.5\times10^7$\,M$_\odot$ for younger ages
and $3.5\times10^7$\,M$_\odot$ for older star formation.
There is no set of parameters which yield a significantly lower mass.
Given the uncertainties involved, the estimates of the stellar light
fraction, the stellar types, the star formation age, and the mass of
the cluster fit together in a remarkably consistent way.
We conclude that at least half, and probably most, of the mass within
0.1\arcsec\ of the nucleus of NGC\,7469 is due to stars rather than
gas.


Using the same star formation models, one finds that the bolometric
luminosity for the nuclear cluster is 1--2$\times10^{10}$\,L$_\odot$.
\cite{gen95} estimated the nuclear bolometric luminosity to be
$1.5\times10^{11}$\,L$_\odot$.
Hence, star formation accounts for 10\% of the nuclear luminosity at
radii small than 0.1\arcsec.

\section{Conclusions}
\label{sec:conc}

We have presented 0.7\arcsec\ radio CO\,2-1 observations and
0.085\arcsec\ adaptive optics K-band spectroscopy of the
central region of NGC\,7469.
Using these data we have investigated the distribution and kinematics of
the cold and warm molecular gas across nearly 2 orders of magnitude in
spatial scale.
Additionally, we have studied the nuclear stellar cluster.

The kinematics of the CO\,2-1 and 1-0\,S(1) lines can be reproduced by
an axisymmetric mass model consisting of a broad disk component, a
circumnuclear ring at 2.3\arcsec, and a nuclear ring at 0.2\arcsec.
The CO\,2-1 morphology also suggests that there may be a
bar or pair of spiral arms between the two rings, although there is no
kinematical signature in the data.
This may be because NGC\,7469 represents the case where straight
shocks cannot form along an inner bar, and hence there is no gas
inflow.
Nevertheless, the increase in gas density in the bar/spiral may have
triggered star formation and this is now seen as the knots of
emission -- which lie inside the radius of the molecular ring.

Comparison of the dynamical and stellar masses indicates that
molecular gas makes up more than half of the total mass within a
radius of 1\,kpc, and that the remaining gas mass is at least several
times that of recently formed stars in the ring.
The CO-to-H$_2$ conversion factor is 0.4--0.8 times the Galactic
conversion factor.
Similar values have been seen a number of times in intense star
forming environments.

The profile of the 1-0\,S(1) emission does not trace the gas
distribution, and much of the flux is likely to arise instead from
X-ray and UV irradiation of gas by the AGN.

By mapping the stellar CO absorption and considering the slope of the
continuum, we have directly resolved the nuclear stellar cluster in
NGC\,7469.
It is extended over 30--60\,pc and contributes 20--30\% of the nuclear
K-band continuum, and about 10\% of the nuclear bolometric luminosity.
The stellar mass counts for at least half, and probably more, of the
total mass within 30\,pc (0.1\arcsec) of the nucleus.


\acknowledgments

The authors are grateful to the support of the staff of the IRAM
interferometer on the Plateau de Bure and in Grenoble, and the staff
of the Keck Observatory.
The authors wish to recognize and acknowledge the very significant
cultural role and reverence that the summit of Mauna Kea has 
always had within the indigenous Hawaiian community.  We are most
fortunate to have the opportunity to conduct observations
from this mountain.
The authors extend their thanks to the following people: Jack
Gallimore and Lowell Tacconi-Garman for considerable help with the CO
data, particularly calibrating and modelling the CO\,1-0 data;
Aaron Evans and Rodger Thompson for providing reduced and calibrated
{\em HST} images;
Andrew Baker for helpful discussions about disk dynamics;
the anonymous referee for several suggestions about how to improve the
paper.



\clearpage


\begin{deluxetable}{ll}

\tabletypesize{\small}
\tablecaption{Derived Properties of the Molecular Gas\label{tab:COprop}}
\tablehead{

\colhead{Component} & 
\colhead{Value} \\

}
\startdata

CO\,1-0 flux (2.0\arcsec\ resolution)	& 300\,Jy\kms \\
CO\,2-1 flux (0.7\arcsec\ resolution)	& 470\,Jy\kms \\
Inclination				& 45$^\circ$ \\
Position Angle				& 128$^\circ$ \\
Systemic Velocity			& 4925\kms \\
Dynamical Mass (within $R<5$\arcsec)	& $9\times10^9$\,M$_\odot$ \\

\enddata

\end{deluxetable}


\begin{deluxetable}{ccccc}

\tabletypesize{\small}
\tablecaption{Parameters of the Mass Model\label{tab:massmod}}
\tablehead{

\colhead{Component} & 
\colhead{Radial Offset} & 
\colhead{FWHM} & 
\colhead{Peak} & 
\colhead{Mass} \\

\colhead{} & 
\colhead{\arcsec} & 
\colhead{\arcsec} & 
\colhead{Scaling} & 
\colhead{$10^9$\,M$_\odot$} \\

}
\startdata

disk    & 0.0 & 4.8 & 1.5 & 3.97 \\
ring    & 2.3 & 0.8 & 3.3 & 4.15 \\
nucleus & 0.2 & 0.6 & 9.3 & 0.88 \\

\enddata

\tablecomments{All components are Gaussians, and trace the mass
surface density distributions.
In the model this is the same as the light surface density.}
\end{deluxetable}


\begin{deluxetable}{llrrrr}

\tabletypesize{\small}
\tablecaption{Late type stellar parameters\label{tab:stars}}
\tablehead{

\colhead{starr} & 
\colhead{stellar} & 
\colhead{$M_{\rm V}$} & 
\colhead{$M_{\rm K}$} & 
\colhead{Mass} & 
\colhead{$W_{\rm CO}$} \\

\colhead{} & 
\colhead{type} & 
\colhead{} & 
\colhead{} & 
\colhead{($M_\odot$)} & 
\colhead{(\AA)} \\

}
\startdata

HD 213758 & M1 I     & $-5.6$ & $-9.5$ & 16\phm{.0} & 32.5 \\
HR 8726   & K5 Ib    & $-5.8$ & $-9.3$ & 13\phm{.0} & 23.3 \\
HR 8465   & K1.5 Ib  & $-5.9$ & $-8.3$ & 13\phm{.0} & 21.0 \\
HR 4517   & M1 III   & $-0.5$ & $-4.6$ & 1.2        & 17.9 \\ 
HR 6705   & K5 III   & $-0.2$ & $-3.8$ & 1.2        & 17.3 \\ 
HR 7806   & K2.5 III & $+0.5$ & $-2.1$ & 1.2        & 13.2 \\
HR 8694   & K0 III   & $+0.7$ & $-1.6$ & 1.1        &  8.3 \\
HR 6703   & G8.5 III & $+0.8$ & $-1.4$ & 1.1        &  5.3 \\
HR 8085   & K5 V     & $+7.4$ & $+4.6$ & 0.7        &  5.9  \\
HR 1084   & K2 V     & $+6.4$ & $+4.2$ & 0.7        &  4.0  \\

\enddata

\tablecomments{Cols 1--2 give the names and types of the stellar
templates from \cite{wal97}, 
and col 6 is the 2.3\,$\mu$m CO equivalent width measured from these;
data in cols 3--5 are from \cite{cox00}.}

\end{deluxetable}


\clearpage


\end{document}